\documentclass[twocolumn]{aastex62}

\usepackage{xcolor}
\usepackage{amsmath}

\graphicspath{{./}{figures/}}

\newcommand{\new}[1]{{#1}}

\begin{document}

%TC:ignore
\title{\textbf{The Roasting Marshmallows Program with IGRINS on Gemini South II -- WASP-121 b has super-stellar C/O and refractory-to-volatile ratios}}

\author[0000-0002-9946-5259]{Peter C. B. Smith}
\affiliation{School of Earth $\&$ Space Exploration, Arizona State University, Tempe AZ 85287, USA}

\author[0000-0002-9142-6378]{Jorge A. Sanchez}
\affiliation{School of Earth $\&$ Space Exploration, Arizona State University, Tempe AZ 85287, USA}

\author[0000-0002-2338-476X]{Michael R. Line}
\affiliation{School of Earth $\&$ Space Exploration, Arizona State University, Tempe AZ 85287, USA}

\author[0000-0003-3963-9672]{Emily Rauscher}
\affil{Department of Astronomy and Astrophysics, University of Michigan, Ann Arbor, MI, 48109, USA}

\author[0000-0003-4241-7413]{Megan Weiner Mansfield}
\affiliation{Steward Observatory, University of Arizona, Tucson, AZ, USA}
\affiliation{NHFP Sagan Fellow}

\author[0000-0002-1337-9051]{Eliza M.-R. Kempton}
\affil{Department of Astronomy, University of Maryland, College Park, MD 20742}

\author[0000-0002-2454-768X]{Arjun Savel}
\affil{Department of Astronomy, University of Maryland, College Park, MD 20742}

\author[0000-0003-3191-2486]{Joost P. Wardenier}
\affil{Institut Trottier de Recherche sur les Exoplan` etes, Universit´ e de Montr´ eal, Montr´ eal, Qu´ ebec, H3T 1J4, Canada}

\author[0000-0002-1321-8856]{Lorenzo Pino}
\affil{INAF -- Osservatorio Astrofisico di Arcetri, Largo Enrico Fermi 5, 50125 Firenze, Italy}

\author[0000-0003-4733-6532]{Jacob L.\ Bean}
\affiliation{Department of Astronomy \& Astrophysics, University of Chicago, Chicago, IL, USA}

\author[0000-0002-6980-052X]{Hayley Beltz}
\affil{Department of Astronomy, University of Maryland, College Park, MD 20742}

\author[0000-0002-2513-4465]{Vatsal Panwar}
\affil{Department of Physics, University of Warwick, Coventry CV4 7AL, UK.}
\affil{Centre for Exoplanets and Habitability, University of Warwick, Coventry CV4 7AL, UK.}

\author[0000-0002-7704-0153]{Matteo Brogi}
\affil{Department of Physics, University of Turin, Via Pietro Giuria 1, I-10125, Turin, Italy}
\affil{INAF -- Osservatorio Astrofisico di Torino, Via Osservatorio 20, I-10025, Pino Torinese, Italy}

\author[0000-0003-0217-3880]{Isaac Malsky}
\affil{Department of Astronomy and Astrophysics, University of Michigan, Ann Arbor, MI, 48109, USA}

\author[0000-0002-9843-4354]{Jonathan Fortney}
\affiliation{Department of Astronomy and Astrophysics, University of California (UCSC), High St 1156, Santa Cruz, 95064, CA, USA.}

\author[0000-0002-0875-8401]{Jean-Michel Desert}
\affil{Anton Pannekoek Institute of Astronomy, University of Amsterdam, Amsterdam, Netherlands}

\author[0000-0002-8573-805X]{Stefan Pelletier}
\affiliation{Observatoire astronomique de l'Université de Genève, 51 chemin Pegasi 1290 Versoix, Switzerland}

\author[0000-0001-9521-6258]{Vivien Parmentier}
\affil{Université de la Côte d’Azur, Observatoire de la Côte d’Azur, CNRS, Laboratoire Lagrange, France}

\author[0009-0005-4890-3326]{Krishna Kanumalla}
\affiliation{School of Earth $\&$ Space Exploration, Arizona State University, Tempe AZ 85287, USA}

\author[0000-0003-0156-4564]{Luis Welbanks}
\affil{School of Earth and Space Exploration, Arizona State University, Tempe, AZ 85287}
\affil{NHFP Sagan Fellow}

\author[0000-0003-1227-3084]{Michael Meyer}
\affil{Department of Astronomy and Astrophysics, University of Michigan, Ann Arbor, MI, 48109, USA}

\author[0000-0002-3380-3307]{John Monnier}
\affil{Department of Astronomy and Astrophysics, University of Michigan, Ann Arbor, MI, 48109, USA}

\begin{abstract}
    A primary goal of exoplanet science is to measure the atmospheric composition of gas giants in order to infer their formation and migration histories. Common diagnostics for planet formation are the atmospheric metallicity ([M/H]) and the carbon-to-oxygen (C/O) ratio as measured through transit or emission spectroscopy. The C/O ratio in particular can be used to approximately place a planet's initial formation radius from the stellar host, but a given C/O ratio may not be unique to formation location. This degeneracy can be broken by combining measurements of both the C/O ratio and the atmospheric refractory-to-volatile ratio. We report the measurement of both quantities for the atmosphere of the canonical ultra hot Jupiter WASP-121 b using the high resolution (R=45,000) IGRINS instrument on Gemini South. Probing the planet's direct thermal emission in both pre- and post-secondary eclipse orbital phases, we infer that WASP-121 b has a significantly super-stellar C/O ratio of 0.70$^{+0.07}_{-0.10}$ and a moderately super-stellar refractory-to-volatile ratio at 3.83$^{+3.62}_{-1.67} \times$ stellar. This combination is most consistent with formation between the soot line and H$_2$O snow line, but we cannot rule out formation between the H$_2$O and CO snow lines or beyond the CO snow line. We also measure velocity offsets between H$_2$O, CO, and OH, potentially an effect of chemical inhomogeneity on the planet day side. This study highlights the ability to measure both C/O and refractory-to-volatile ratios via high resolution spectroscopy in the near-infrared H and K bands.

\end{abstract}

% \authorcomment1{author text}


\received{2024 July 5}
\revised{2024 September 30}
\accepted{2024 October 8}
% \published{XXX}
\submitjournal{\aj}

\section{Introduction}

Understanding how solar systems and the planets within them form is a key goal of planetary astronomy. Under the core accretion model of planet formation, gas giant planets are hypothesized to form exterior to the H$_2$O snow line, where temperatures in the protoplanetary disk are low enough for the condensation of volatile-carrying molecules like H$_2$O (``ices") into solids, providing enough solid material for a $\sim$10$M_\oplus$ core to form and accrete a H-He envelope \citep{lewis1972,hayashi1981,pollack1996, hubickyj2005}. The discovery of hot Jupiters, short period (P $<$ 10 day) gas giant planets on close-in orbits around their stars, initially challenged this conventional model of planet formation \citep{mayor1995}. Recent studies have explored the possibility of hot Jupiters forming interior to the H$_2$O snow line \citep{batygin2016,madhu2017}, but it is hypothesized that these planets initially formed exterior to the snow line then migrated to their current orbits \citep{lin1996, fortney2012}.

The radial distance at which a planet initially formed as well as its migration history will determine the overall enrichment and ratios of volatile elements, like C and O, of the material the planet accretes \citep{oberg2011, madhu2014}. Gas interior to the H$_2$O snow line is expected to have a stellar C/O ratio, while once beyond the H$_2$O snow line, the gas phase C/O ratio should increase with distance from the star depending on which snow lines are crossed \citep{oberg2011, oberg2016, schneider2021}. Because hot Jupiters are hot enough for much of their volatile inventory to remain gaseous in molecules like H$_2$O and CO, their atmospheric volatile ratios can be probed via spectroscopy. Thus, a major goal of exoplanet science has been to measure hot Jupiter volatile enrichment and ratios (e.g., [C/H], [O/H], and C/O) and tie these quantities back to potential formation conditions.

Numerous spectroscopic campaigns with the Hubble (HST) and Spitzer Space Telescopes have attempted to measure the volatile content of hot Jupiters (e.g., \citealp{kreidberg2014, line2016,arcangeli2018, mansfield2021}). However, reliable measurements of the C/O ratio were challenging to acquire as HST/WFC3 primarily probed a broadband H$_2$O feature and observations with Spitzer were usually limited to 2-4 photometric filters and thus often ill suited for precise compositional inference. While precise measurements of CO and CO$_2$ were lacking, a potential trend of subsolar H$_2$O abundances began to emerge among the hot Jupiter population \citep{pinhas2018, welbanks2019}. Theoretical advancements in ground-based high resolution spectroscopy \citep{brogi2019, gibson2020} and the launch of JWST have enabled a handful of hot Jupiter C/O measurements in recent years (e.g., \citealp{line2021, pelletier2021, bean2023, taylor2023, august2023}). Just as with the earlier HST/WFC3 campaigns, there is an emerging trend of hot Jupiters with subsolar H$_2$O abundances paired with supersolar CO abundances and thus high (0.8-1, compared to the solar value of 0.55, \citealp{asplund2009}) C/O ratios \citep{pelletier2021, boucher2021, lesjak2023}.

Given the inferred super-stellar C/O ratios, many of these planets have been interpreted to have formed beyond the H$_2$O snow line. However, recent studies have identified pathways in which a giant planet could form interior to the H$_2$O snow line yet have a final gas phase C/O that is super-stellar \citep{lothringer2021, chachan2023}. Such planets could have instead formed between the ``soot line" -- the distance at which refractory carbon sublimates -- and the H$_2$O snow line. In this case, a large portion of the atmospheric O would be sequestered into silicates, hence the measured subsolar H$_2$O abundances and high C/O ratios.

Measured high C/O ratios for hot Jupiters are then degenerate between formation interior or exterior to the H$_2$O snow line. This degeneracy can be broken by measuring both a planet's volatile and refractory (elements with high condensation temperatures; e.g., Fe, Mg, Si, \citealp{lodders2003}) content because their relative enrichments will be more unique to specific formation pathways \citep{schneider2021, lothringer2021, chachan2023}. Planets formed via the new proposed pathway between the soot and snow lines will have both a superstellar C/O and refractory-to-volatile ratio, while planets that formed beyond the snow line will have low refractory-to-volatile ratios. However, much of the refractory content of planets with $T_\mathrm{eq} \lesssim 2000$ K  is condensed out of the atmosphere and not accessible via spectroscopy. Ultra-hot Jupiters (UHJs), on the other hand, with their high equilibrium temperatures ($T_\mathrm{eq} \gtrsim 2200$ K), present both volatile and refractory species in the gas phase, hence enabling the abundances of both to be determined spectroscopically \citep{lothringer2018, arcangeli2018, bell2018}. UHJs are thus excellent targets for breaking the C/O-formation degeneracy and testing the hypothesized formation scenario between the soot line and H$_2$O snow line.

WASP-121 b is perhaps one of the most extensively studied UHJs. It was one of the prototypical examples, providing the first confirmed thermal inversion in a transiting exoplanet \citep{mikalevans2017}. WASP-121 b has been studied both in transmission and emission from ground- and space-based observatories \citep{gibson2020, hoeijmakers2020,merritt2020, wilson2021, maguire2023, hoeijmakers2024}. \cite{changeat2024} recently measured WASP-121 b via a HST/WFC3 phase curve, from which they inferred a supersolar C/O ratio and conclude the planet formed beyond the snow line. \cite{lothringer2021} used combined HST measurements with STIS and WFC3 to infer a super-solar refractory-to-volatile ratio. However, the volatile measurements were limited to H$_2$O, and they note that without measurements of CO, they lacked a complete grasp of the total O inventory and any grasp of the C inventory. With an enriched refractory-to-oxygen ratio, WASP-121 b's atmospheric composition may be consistent with the proposed formation scenario between the soot and snow lines, but uniform measurements of all three of the carbon, oxygen, and refractory inventories are needed to rule out other formation pathways.

Refractory species are typically more readily accessible in the optical and volatile species in the IR, necessitating the use of multiple instruments to measure the refractory-to-volatile ratio \citep{lothringer2021, kasper2023}.  However, recently, Fe I was detected in the atmosphere of UHJ MASCARA-1 b via K band measurements with the CRIRES+ instrument on the VLT \citep{ramkumar2023}. This study demonstrates that both refractory and volatile species can be probed simultaneously with current ground-based NIR instruments, potentially including IGRINS on Gemini-South, which has simultaneous H and K band coverage at R$\approx$45,000. The simultaneous measurement of both refractories and volatiles with a single instrument forgoes biases that could arise from e.g., differences in data reduction methods or other systematic offsets.

In this work, we present an analysis of IGRINS pre- and post-eclipse observations of the direct thermal emission of WASP-121 b in order to measure both its volatile and refractory content, enabling the inference of diagnostic elemental ratios needed to assess the formation histories of these enigmatic worlds. The observations and data reduction are detailed in Section \ref{sec:obs}. In Sections \ref{sec: cross correlation} we describe the application of cross-correlation analyses to detect WASP-121 b's atmosphere and the individual gases within it, and in Section \ref{sec: dynamics} we detail the search for signs of atmospheric dynamics. The application of atmospheric retrieval techniques for measurement of the planet's composition and vertical thermal structure is discussed in Section \ref{sec: retrieval}, and the implications of these measurements are discussed in Section \ref{sec: discussion}. The paper is summarized and concluded in Section \ref{sec:conclusions}.

\section{Observations and Data Reduction}
\label{sec:obs}

\begin{table}[]
    \centering
    \begin{tabular}{c c c}
     \hline
     \hline
     Name & Value & Reference \\
     \hline
       & Stellar Parameters   &  \\
        $R_\mathrm{star}$ & 1.44 [$R_\odot$] & \cite{borsa2021} \\
        $M_\mathrm{star}$ & 1.38 [$M_\odot$] & \cite{borsa2021} \\
        $T_\mathrm{eff}$ & 6586 $\pm$ 59 [K] & \cite{borsa2021} \\
        $K$ mag. & 9.347 $\pm$ 0.022 & \cite{2MASS} \\
        $\gamma$ & 38.198 $\pm$ 0.002 [km s$^{-1}$] & \cite{borsa2021} \\
        $\mathrm{[O/H]}$ & 0.42 $\pm$ 0.07 & \cite{polanski2022} \\
        $\mathrm{[C/H]}$ & 0.04 $\pm$ 0.05 & '' \\
        $\mathrm{[Fe/H]}$ & 0.24 $\pm$ 0.03 & '' \\
        $\mathrm{[Mg/H]}$ & 0.15 $\pm$ 0.04 & '' \\
        $\mathrm{[Ca/H]}$ & 0.25 $\pm$ 0.03 & ''\\
        $\mathrm{[Si/H]}$ & 0.24 $\pm$ 0.03 & ''\\
        $\mathrm{[V/H]}$ & 0.01 $\pm$ 0.06 & ''\\
        $\mathrm{[Ti/H]}$ & 0.22 $\pm$ 0.04& ''\\
        $\mathrm{[Cr/H]}$ & 0.23 $\pm$ 0.04 & ''\\
        C/O & 0.23 $\pm$ 0.05 & '' \\
        & & \\
        \hline
        & Planet Parameters & \\
        $R_P$ &  1.865 $\pm$ 0.044 [$R_\mathrm{Jup}$] & \cite{delrez2016} \\
        $M_P$ &  1.183$^{+0.064}_{-0.062}$ [$M_\mathrm{Jup}$] & \cite{delrez2016} \\
        $T_\mathrm{eq}$ & 2358 $\pm$ 52 [K] & \cite{delrez2016} \\
        $T_0$ & 2458119.72074  & \cite{bourrier2020} \\
        & $\pm$ 0.00017 [BJD] & \\
        $\Delta T_0$ & -28$^{+18}_{-17}$ [s] & \cite{mikalevans2023} \\
        $P$ & 1.27492504 & \cite{bourrier2020} \\
         & $\pm$ 1.5$\times 10^{-7}$ [day] & \\
        $a$ & 0.02544$^{+0.00049}_{-0.00050}$ [A.U.] & \cite{delrez2016} \\
        $K_P$ & 217.08 $\pm$ 4.27 [km s$^{-1}$] &  Derived \\
        $v\sin i$ & 7.15 $\pm$ 0.18 [km s$^{-1}$] & Derived \\
        \hline 
        \hline
        
    \end{tabular}
    \caption{Relevant system parameters and their references.}
    \label{tab:params}
\end{table}

\begin{figure*}
    \centering
    \includegraphics[width=0.7\textwidth]{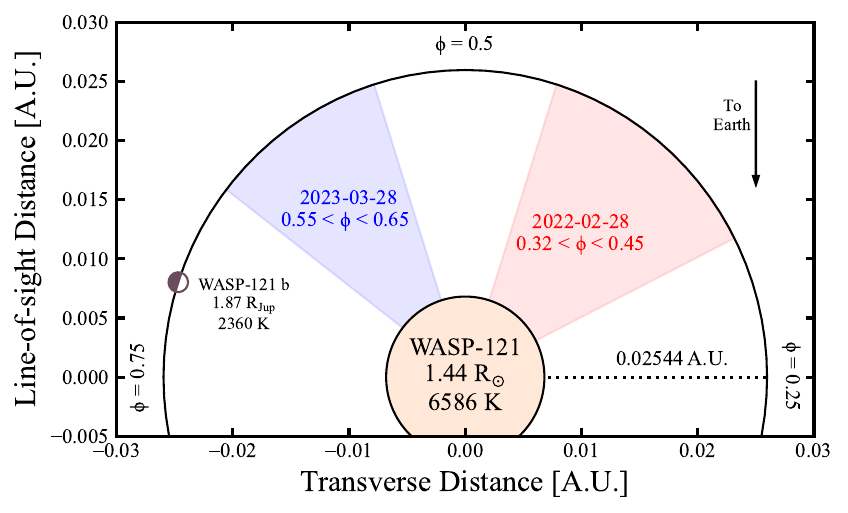}
    \includegraphics[width=\textwidth]{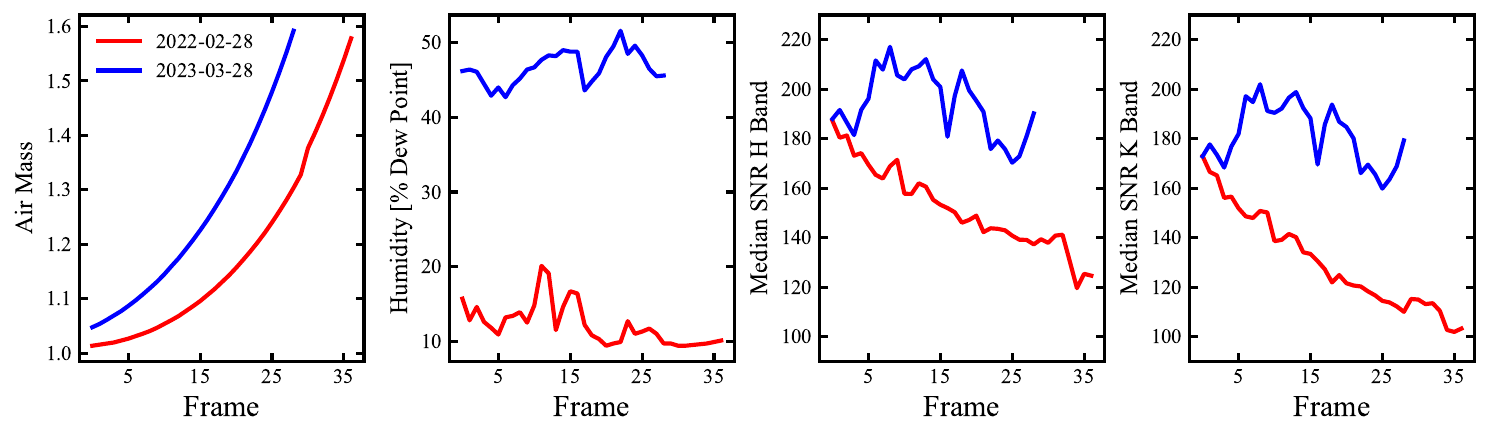}
    \caption{\textit{Top}: To-scale schematic of the WASP-121 system from a top-down perspective. \textit{Bottom}: Observing conditions during each of the two sequences.}
    \label{fig:schematic}
\end{figure*}

We present two separate nights of data capturing the direct thermal emission of WASP-121 b using IGRINS \citep{park2014, mace2016} on Gemini South\footnote{The reduced data products as well as the model spectra used in this paper are publicly available in a Zenodo repository: \href{https://doi.org/10.5281/zenodo.12635249}{https://doi.org/10.5281/zenodo.12635249}}. The first sequence was taken on UTC February 28, 2022 as part of the Large-and-Long Program ``Roasting Marshmallows: Disentangling Composition \& Climate in Hot Jupiter Atmospheres through High-Resolution Thermal Emission Cross-Correlation Spectroscopy" (GS-2022-LP-206, PI M. Line). It consisted of a continuous 4.2 hour sequence of 150s exposures in an AB-BA nodding pattern while the planet was in the pre-secondary eclipse phases (0.32 $<$ $\phi$ $<$ 0.45, 37 AB pairs). The median SNR in the H and K bands were 155 and 130, respectively, and we will hereafter refer to this as the ''pre-eclipse" sequence. 

The second sequence was taken on UTC March 28, 2023 as part of the queue program ''Tracing the Day-Night Structure of WASP-121b with Multi-Phase High-Resolution Spectroscopy" (GS-2023A-Q-222, PI E. Rauscher). It consisted of a 3 hour sequence of 150s exposures and was taken in the post-eclipse phases (0.55 $<$ $\phi$ $<$ 0.65, 29 AB pairs). The median H and K band SNR's were 200 and 180 and we will hereafter refer to this as the ''post-eclipse" sequence. The observing conditions and phase coverage during each sequence are summarized in Figure \ref{fig:schematic}.

For each sequence, the raw data were calibrated and 1D spectra were extracted by the IGRINS team using the IGRINS Pipeline Package (\citealp[PLP;][]{lee1016, mace2016}) per AB pair, hereafter referred to as frames. We then organized the data into cubes of shape $N_\mathrm{order} \times N_\mathrm{frame} \times N_\mathrm{pixel}$ using the same procedures described in \cite{line2021} and \cite{brogi2023}. This includes an adjustment to the PLP wavelength solution for every frame, discarding orders with heavy telluric contamination, and trimming 200 low throughput pixels at the edges of each order. Our Python routines for processing and organizing the PLP output, as well as extracting relevant information from the FITS headers and detrending the data, are publicly available on GitHub\footnote{\href{https://github.com/petercbsmith/cubify}{https://github.com/petercbsmith/cubify}}.

To detrend the data, we apply a singular value decomposition (SVD) to identify and remove the first few singular vectors in each order \citep{dekok2013, line2021, brogi2023}. Operationally, this is done by decomposing the $N_\mathrm{frame} \times N_\mathrm{pixel}$ data matrix then setting the first few singular values to zero before recomposing the spectral matrix, effectively subtracting the first few right singular vectors. We tested different numbers of vectors to remove and found no significant differences in cross-correlation maps or velocity inferences (described in the following two sections) between removing the first 3, 4, 5, or 6 vectors. This robustness against number of vectors removed may be partially due to the relatively constant humidity during each observed sequence. Variable humidity has been shown to make high resolution data more sensitive to the number of singular vectors or principal components removed \citep{smith2024}. For the analyses described in this paper, we choose to remove the first 4 for both sequences because this is enough to remove any visual telluric features in most orders. We save two matrices of shape $N_\mathrm{frame} \times N_\mathrm{pixel}$ per order: one in which the first 4 singular vectors have been removed and one in which all \textit{but} the first 4 have been removed for model processing \citep{brogi2019, line2021}. To calculate the orbital phase at each frame as well as the planet's expected radial velocity semi-amplitude, we assume a circular orbit and use the period and semi-major axis reported in \cite{bourrier2020} and the updated mid-transit time from \cite{mikalevans2023}.

\section{Molecular Detection via Cross-Correlation}
\label{sec: cross correlation}

\begin{figure*}
    \centering
    \includegraphics[width=\linewidth]{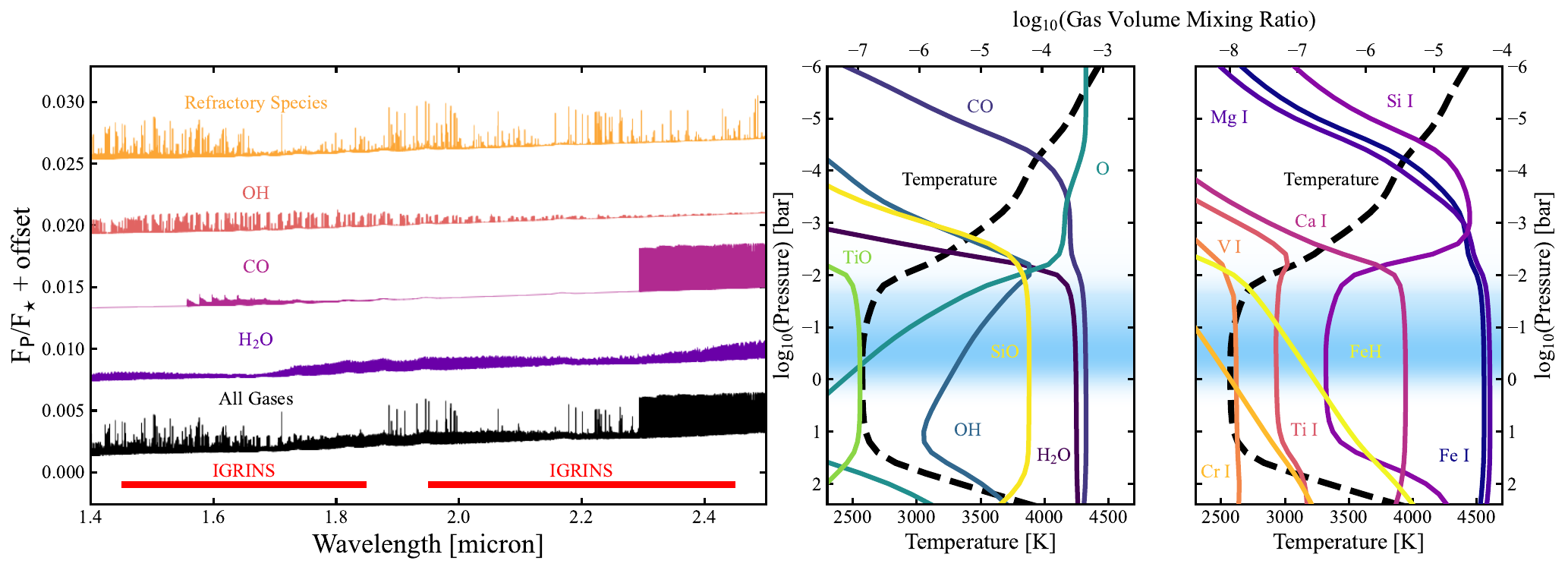}
    \caption{Expectations for WASP-121 b's atmosphere via solar composition 1D radiative-convective-thermoequilibrium (1D-RCTE). \textit{Left}: Output high resolution spectra from the solar composition 1D-RCTE model as described in Section \ref{sec: cross correlation}, in units of stellar contrast. A model spectrum with all expected opacity sources over the IGRINS wavelength range is shown in black, while the colored spectra contain subsets of gases. \textit{Right}: The output vertical thermal structure and gas volume mixing ratios as predicted by the 1D-RCTE model. Also shown is the wavelength integrated contribution function (blue shading), indicating which pressures IGRINS is most sensitive to. These pressures include thermal dissociation of several O-bearing molecules and ionization of atomic gases, indicating WASP-121 b may have muted spectral features. As these gases dissociate, much of the total O inventory transfers to atomic O, which does not have any spectral features IGRINS is sensitive to. This can challenge efforts to estimate the total volatile inventory of UHJs (see e.g., \citealp{brogi2023}).}
    \label{fig:solarcomp}
\end{figure*}

As an initial analysis to assess the strength of the planetary signal and to identify potential sources of opacity in the planet's outgoing spectrum, we calculate 2D cross-correlation maps by cross-correlating a model spectral template with the data (e.g., \citealp{dekok2013, birkby2013, brogi2016}). We use the \texttt{ScCHIMERA} framework to calculate a solar composition ([M/H] = 0; C/O = 0.55) 1D radiative-convective-thermoequilibrium (1D-RCTE) model as described in \cite{arcangeli2018, piskorz2018} and \cite{weinermansfield2018}. The heat redistribution factor, $f$\footnote{$f = (T_\mathrm{day}/T_\mathrm{eq})^4$, \citep{fortney2005}}, is set to 2.2 following the trend with equilibrium temperature predicted by \cite{parmentier2021}. \texttt{ScCHIMERA} provides day side averaged pressure-temperature (P-T) and gas volume mixing ratio (VMR) profiles which we pass through a GPU-accelerated version of CHIMERA \citep{line2013, line2021} to calculate a R=250,000 emission spectrum. We include continuum opacities\footnote{CIA from \cite{karman2019}; CO and OH from HITEMP \citep{li2015, rothman2010}; H$_2$O from \cite{polyansky2018}; FeH from \cite{bernath2020}; VO from \cite{mckemmish2016}; TiO from \cite{mckemmish2019}; SiO from \cite{barton2013}; and the atomics from \cite{kurucz2018}. Atomic species from the Kurucz line database.  Cross-sections for H$_2$O, FeH, VO, and TiO were generated as described in \cite{gharib-nezhad2021}; for CO and OH were generated with HELIOS-K \citep{grimm2015,grimm2021}; for H$^-$ generated as described in \cite{john1988}.} from H$_2$-H$_2$ and H$_2$-He collision induced absorption, H$^{-}$ bound-free continuum, and H-e$^{-}$ free-free continuum; opacities of the main volatile-carrying gases H$_2$O, $^{12}$CO, and OH; and the opacities of the following refractory-bearing species that have strong lines in the H and K bands: Fe I, Mg I, Ti I, Ca I, Cr I, V I, TiO, VO, SiO, and FeH. The P-T and selected gas VMR profiles are shown in Figure \ref{fig:solarcomp} along with the output high resolution spectrum. Also shown are contributions to the total spectrum from H$_2$O, CO, and OH individually and all of the refractory-bearing species. Before cross-correlation, the model spectrum is convolved with a Gaussian instrumental profile at the nominal resolving power of IGRINS and a equatorial rotation kernel set to the appropriate rotation speed assuming tidal locking ($v\sin i = 7.15$ km s$^{-1}$). \new{To scale the planet spectrum down to units of contrast relative to the stellar continuum, $F_P / F_\star$, we divide the model planet spectrum by an interpolated PHOENIX model stellar spectrum \citep{husser2013} at the appropriate $T_\mathrm{eff}$ and log$g$ and smoothed with a Gaussian kernel with a standard deviation of 250 elements. The effect of phase-dependent line depths in the planet-star contrast spectrum is reproduced by injecting this scaled planet spectrum into the data as described below.}
 
\begin{figure}
    \centering
    \includegraphics[width=\linewidth]{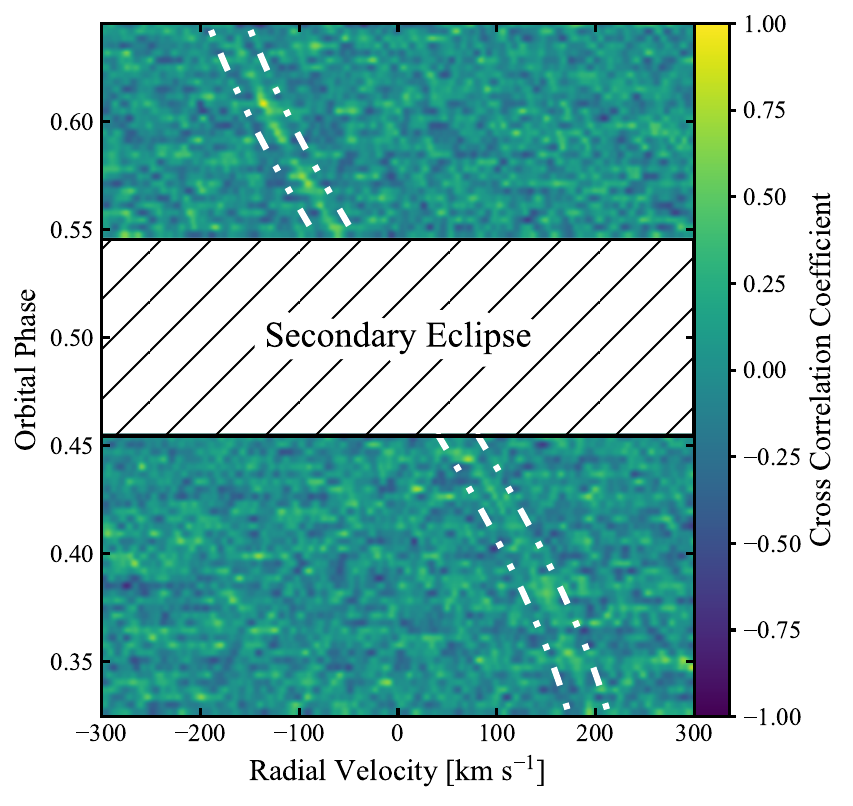}
    \caption{Cross-Correlation ``trail" as described in Section \ref{sec: cross correlation}. The solar composition 1D-RCTE model spectrum was cross-correlated with the post-SVD at each frame in both observational sequences along on a grid of possible line-of-sight radial velocities in the system barycentric frame. The dotted white lines indicate the planet signal's expected path in velocity-time space offset by $\pm$ 20 km s$^{-1}$ for clarity. The hatched region indicates secondary eclipse, during which the planet is occulted by the star and data were not taken.}
    \label{fig:trail}
\end{figure}

The true planet signal is constantly Doppler shifting over the course of each observed sequence, and at any time $t$ the planet's line-of-sight velocity can be described by:

\begin{equation}
    V_\mathrm{LOS}(t) = \gamma + V_\mathrm{bary}(t)  + K_\mathrm{P} \sin \big[ 2 \pi \phi(t) \big] + dV_\mathrm{sys}
    \label{eqn:velocity}
\end{equation}
where $\gamma$ is the star-planet system's radial velocity, $V_\mathrm{bary}$ is the Solar System barycentric radial velocity in the observatory's rest frame, $K_\mathrm{P}$ is the planet's radial velocity semiamplitude in the star-planet barycentric frame, $\phi$ is the orbital phase, and $dV_\mathrm{sys}$ is an additive term to account for any systematic offset. To search for this characteristic motion, we cross-correlate the solar composition 1D-RCTE model spectrum with the post-SVD data at each frame along a grid of possible line-of-sight radial velocities in the stellar rest frame. To reproduce any alterations to the true planet signal by the detrending process, we Doppler shift the model \new{planet spectrum} to the test velocity \new{before dividing by the smoothed model stellar spectrum,} then inject the \new{scaled and} Doppler shifted model spectrum into the scaling matrix with the higher-order singular vectors removed. The first four singular vectors \new{are then removed} via SVD again before cross-correlation. \new{This injection process has the benefit of both altering the model spectrum in a similar manner as the true underlying planet signal has been via the SVD, as well as converting this model to $F_P / F_\star$ in a fashion that is largely independent of the stellar model beyond the baseline continuum level.} The resultant cross-correlation ``trail" is shown in Figure \ref{fig:trail}, where a faint signal along the expected path from Equation \ref{eqn:velocity} can be seen.

\begin{figure*}
    \centering
    \includegraphics[width=\textwidth]{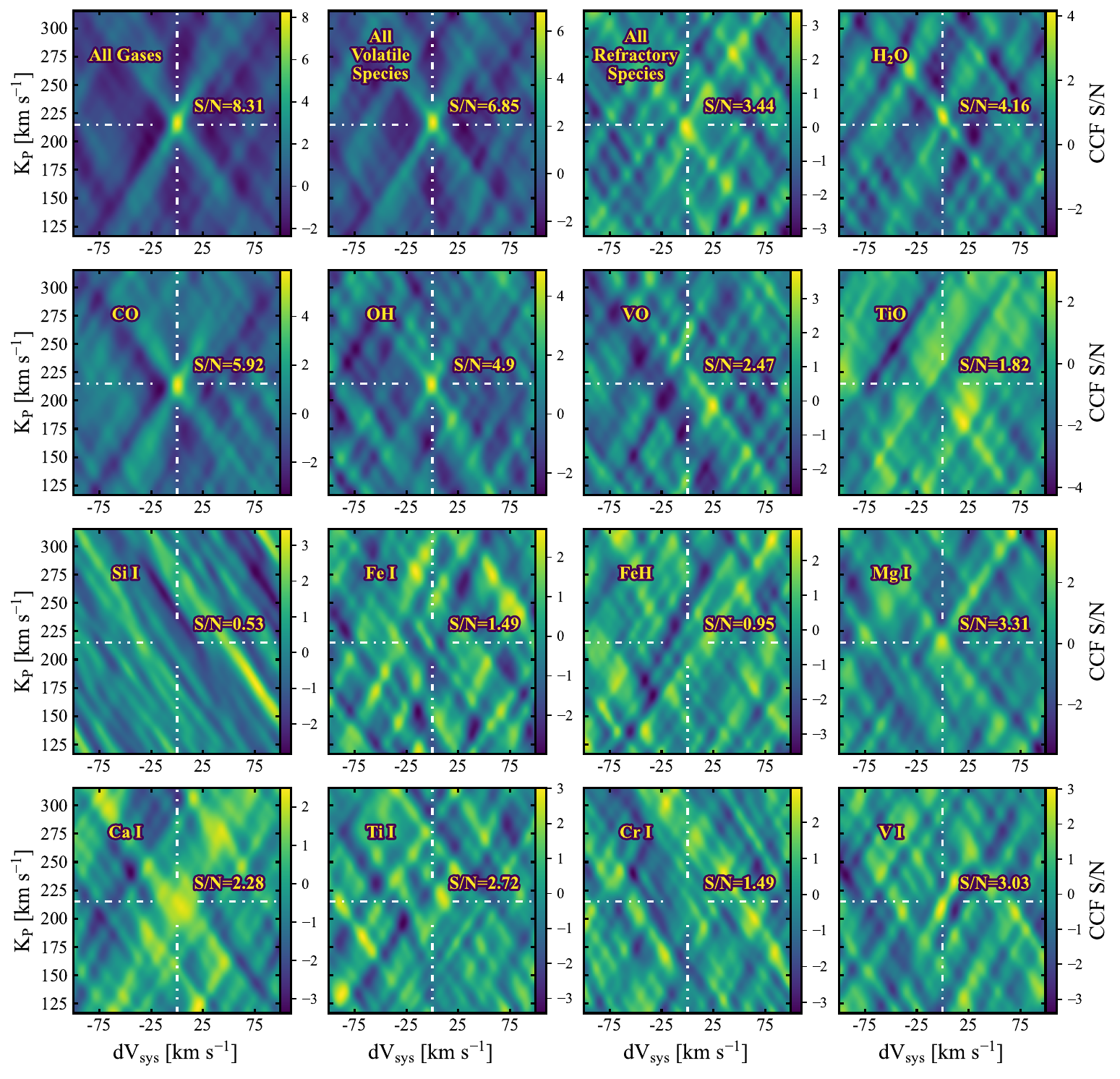}
    \caption{Cross-correlation detection maps calculated using the methods and models described in Section \ref{sec: cross correlation}. The ``All Gases" map in the top left is the result of cross-correlating the solar composition 1D-RCTE model spectrum including all sources of opacity listed in that same section. Every other panel displays the residuals between the ``All Gases" map and the resultant CCF map from cross-correlating a model with all sources of opacity \textit{except} the gas listed in the upper left of each individual panel. The dashed white lines indicate the expected $K_P$ and $dV_\mathrm{sys}$ of the true planet signal, and the quoted S/N values are the maximum value found within a 20 km s$^{-1} \times$ 20 km s$^{-1}$ box centered at the expected $K_P$ and $dV_\mathrm{sys}$.}
    \label{fig:CCF maps}
\end{figure*}

To build signal, we cross-correlate the model spectrum with the post-SVD data again, this time along a grid of possible values for $K_\mathrm{P}$ and $dV_\mathrm{sys}$ and summing over all frames, yielding a 2D cross-correlation function (CCF) map. We then median subtract and normalize the map by the 3-$\sigma$ clipped standard deviation to obtain the cross-correlation signal-to-noise (CCF S/N, \citealp{kasper2021, kasper2023}). Using the solar composition 1D-RCTE model, we detect WASP-121 b's atmosphere with a CCF S/N of 8.31 (Figure \ref{fig:CCF maps}, top left). It should be noted that the model P-T profile has a thermal inversion in the infrared photosphere, resulting in molecular and atomic emission lines rather than absorption. The planet cross-correlation signal is positive, indicating that we are indeed detecting emission features. This confirms the presence of a thermal inversion as inferred by previous studies both at low \citep{mikalevans2017, mikalevans2020, changeat2024} and high \citep{hoeijmakers2024} spectral resolution.

\begin{figure*}
    \centering
    \includegraphics[width=\textwidth]{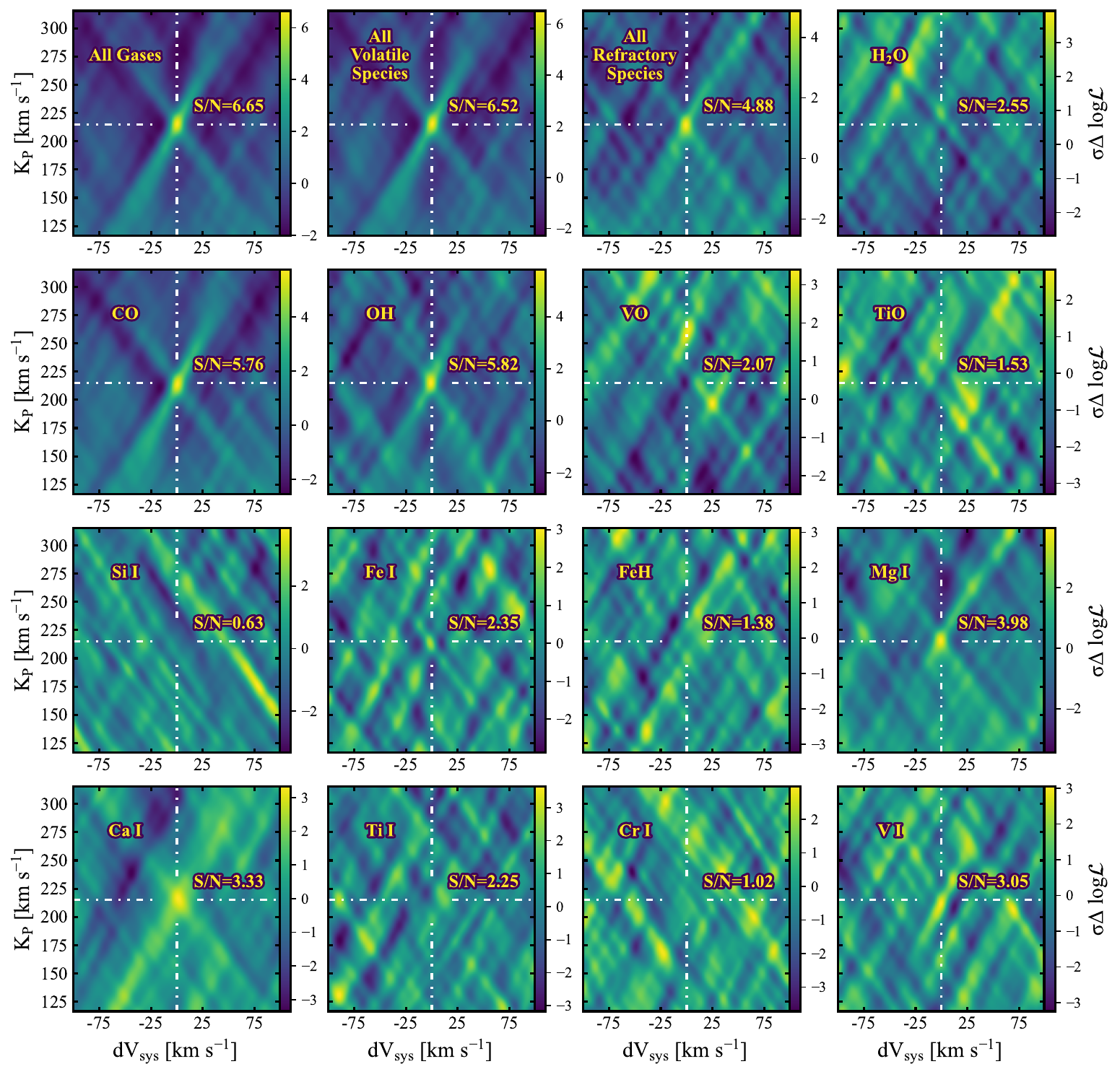}
    \caption{Similar to Figure \ref{fig:CCF maps} but using the log-likelihood function from \cite{brogi2019} instead of cross-correlation coefficients.}
    \label{fig:logL maps}
\end{figure*}

To search for individual gases, it is common in the literature to create ``one-gas-only" spectra by using the same atmospheric structure as a more comprehensive model but removing all opacity sources except for a particular gas of interest before recomputing a spectral template and cross-correlating with the data again (e.g., \citealp{line2021, brogi2023}). However, this method may not accurately capture the true individual line strengths and contrasts by failing to account for the cumulative opacity contributed by the line wings (or even line cores) from the other gases. To isolate the contribution an individual gas has to the total CCF S/N, we recalculate the 1D-RCTE spectral template including all sources of opacity \textit{except} the individual gas of interest and create a new CCF map with this model spectrum. We then take the difference between this new CCF map and the original map created using the model with all sources of opacity and normalize this residual map as described above to estimate the individual gas's CCF S/N.

We make CCF detection maps for H$_2$O, CO, and OH because they will be the primary carriers of WASP-121 b's volatile content and thus are of interest to identify and measure. The normalized residual CCF maps are also shown in Figure \ref{fig:CCF maps}. Using this method, we detect CO emission lines at S/N = 5.92 and OH emission lines at S/N = 4.90. H$_2$O has a weaker detection at S/N = 4.16, and we also note the presence of noise and/or aliasing structure in the CCF map that is of similar amplitude to the suspected true signal. The CCF peak for H$_2$O is also visually at a higher $K_P$ than CO and OH (see more quantitative estimates on $K_P$ and $dV_\mathrm{sys}$ in the following section). Despite the much more numerous H$_2$O lines in the H and K bands compared to CO and OH, this weak H$_2$O detection is not unexpected as previous studies with HST found muted H$_2$O features \citep{mikalevans2022, mansfield2021}. Per the 1D-RCTE model, this is expected due to thermal dissociation and a sharp drop off in the H$_2$O volume mixing ratio in the infrared photosphere. The detection of OH also provides further indirect evidence of the thermal dissociation of H$_2$O.

Motivated by recent ground-based detections of refractory species in the infrared \citep{ramkumar2023,parker2024}, we also individually search for the various refractory species included in the full atmospheric template. The residual CCF maps from excluding each individual gas from the model spectrum are also shown in Figure \ref{fig:CCF maps}. Unfortunately, this does not yield strong detections of any individual refractory-carrying gas. Even when removing all of them, their total contribution is only tentatively detected at S/N = 3.44 (``All Refractory Species" panel). However, if we create a similar detection map using only the volatile species ("All Volatile Species" panel), the detection is weaker than the ``All Gases" model at S/N=6.85 (compared to 8.31), further indicating that we are indeed sensitive to the refractory species.

In addition to CCF detection maps, we also calculated $K_P$ and dV$_\mathrm{sys}$ maps using the likelihood formalism from \cite{brogi2019} and normalized these maps and extracted S/N estimates in the same manner as the CCF maps (Figure \ref{fig:logL maps}). We expect slightly better individual gas detecting power because the log-likelihood function (log$\mathcal{L}$) is more sensitive to line shapes and amplitudes than the cross-correlation coefficient, and it is thus slightly less susceptible to aliasing with other gases and noise. Indeed, in the log$\mathcal{L}$ maps, the detections of CO and OH are stronger (S/N = 5.76 and 5.82, respectively) as well as the detection of the ``All Refractory Species" model, which increased to S/N = 4.88. The log$\mathcal{L}$ maps also yield tentative detections of Mg I (S/N=3.98), Ca I (3.33), and V I (3.05). There is also a weak Fe I signal at the expected planet velocities, but it is not stronger than the noise structure in the map. Curiously, the H$_2$O signal strength is decreased compared to the CCF map, to S/N = 2.55. This may indicate a line-amplitude mismatch between the true H$_2$O signal and the solar composition model. Indeed, \cite{changeat2024} recently measured WASP-121 b to have a supersolar C/O ratio, and we confirm this in Section \ref{sec: retrieval}.

We find few differences between the residual ``leave a gas out" method employed here and the more traditional ``one-gas-only" model cross-correlation method that is common in the literature. For comparison, we recalculated each CCF and log$\mathcal{L}$ map and include them in the appendix (Figures \ref{fig:vanilla ccf} and \ref{fig:vanilla logL}). The same gases are detected between both methods, however the H$_2$O and all-refractories models are more robustly detected (CCF S/N $>$ 5) via the more traditional method. The source of this discrepancy is unclear as our weak detections of both using our new proposed method are not unexpected. Future work could be done to simulate similar observations and test whether one method or the other is over- or underestimating detection significance, but that is beyond the scope of this paper.

\section{Searching for Signatures of Atmospheric Dynamics}
\label{sec: dynamics}

Atmospheric dynamics and inhomogeneity can manifest as anomalous Doppler shifts outside expectations from the planet's orbital motion (e.g., \citealp{zhang2017,beltz2022, hoeijmakers2024}). Previous studies in emission have tied velocity asymmetries between pre- and post-eclipse sequences \citep{pino2022} and differences between individual gases \citep{cont2021,brogi2023} to atmospheric dynamics and thermochemical inhomogeneity. To search for both, we use the log-likelihood formalism from \cite{brogi2019} and the nested sampler Pymultinest \citep{feroz2009, buchner2016} to estimate the value of $K_P$ and $dV_\mathrm{sys}$ for WASP-121 b. We do this with the ``All Gases" atmospheric template described in Section \ref{sec: cross correlation} as well as spectral templates including the opacity of only one of each of the more robustly detected gases (H$_2$O, CO, and OH) plus continuum opacity. We also include a multiplicative scaling factor, $a$, as a nuisance parameter to account for line amplitude mismatches as our solar composition model may not be representative of the true atmospheric composition.

Ideally, we would follow the same philosophy as the previous section and avoid inaccurate line strength estimates due to the absence of a cumulative opacity from other gases. However, the path to avoid this is unclear, and our goal here is not to assess the signal strength but only line \textit{positions}. Biases in velocity inferences that can arise from ``one gas only" models are underexplored, and we leave this assessment to future work, but they are likely minimal at the spectral resolving power of IGRINS.

When using the full atmospheric template, we measure WASP-121 b's orbital velocity to be 215.28$^{+0.35}_{-0.34}$ km s$^{-1}$, well within the uncertainty based on the assumption of a circular orbit and the literature reported semimajor axis and period, and also consistent with previous work \citep{hoeijmakers2024}. We also measure a small net red shift with $dV_\mathrm{sys}$ = 1.20$^{+0.13}_{-0.11}$ km s$^{-1}$. This would be consistent with day-to-night winds as viewed from the day side, although these would be expected at lower pressures than we are probing here \citep{kempton2012}. Alternatively, such a red shift could arise from a combination of planetary rotation and an eastward offset hot spot \citep{zhang2017}, although this would be difficult to quantify as hot spot regions are typically more isothermal and thus contribute little Doppler shifting of the total atmospheric signal \citep{vansluijs2023}.

\begin{figure}
    \centering
    \includegraphics[width=\linewidth]{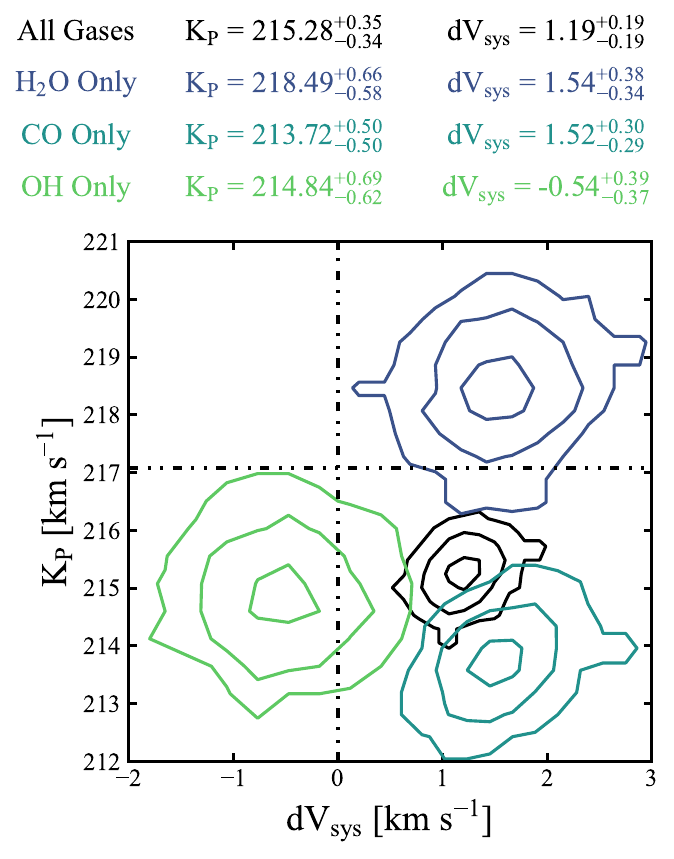}
    \caption{2D posterior distributions for our inferences of WASP-121 b's radial velocity semiamplitude, K$_\mathrm{P}$ and a systematic velocity offset, dV$_\mathrm{sys}$, using four different 1D-RCTE models. The colored posterior distributions used models that included only continuum opacity sources and the opacity of that one particular gas using the P-T and abundance profile as output by the solar composition 1D-RCTE model, while the black posterior distribution is from using the 1D-RCTE model including all gases mentioned in Section \ref{sec: cross correlation}. There is a clear discrepancy in K$_\mathrm{P}$ and dV$_\mathrm{sys}$ between H$_2$O, CO, and OH. H$_2$O appears to have a larger K$_\mathrm{P}$ compared to the other gases, while OH appears to have net relative blueshift by about 1.5 km s$^{-1}$ compared to the other gases. The contours show 1, 2, and 3 $\sigma$ confidence intervals.}
    \label{fig: KpVsys}
\end{figure}

Between the pre- and post-eclipse sequences, our inferred values for $K_P$ and $dV_\mathrm{sys}$ are consistent with each other when using the ``all gases" atmospheric template, and we find no evidence for ephemeris error or a strong equatorial jet when attempting to fit for these quantities such as in \cite{smith2024}. The latter is consistent with previous studies that have measured WASP-121 b to have a small hot spot offset, indicative of strong atmospheric drag \citep{bourrier2020,mikalevans2023, changeat2024, showman2011}. However, when we repeat the $K_P$-$dV_\mathrm{sys}$ inference using the ``one gas only" spectral templates, we find that these quantities are inconsistent between each of the three main volatile-carrying gases. Figure \ref{fig: KpVsys} shows the 2D posterior probability distribution for $K_P$ and $dV_\mathrm{sys}$ from the four different spectral templates and combining the pre- and post-eclipse sequences. Notably, CO and OH have similar effective $K_P$'s while H$_2$O appears to have a larger effective $K_P$ by $\sim$ 4 km s$^{-1}$. Likewise, the H$_2$O and CO models yield consistent estimates on $dV_\mathrm{sys}$, while the value inferred from the OH model is much lower. The inferred values using the ``all gases" model appears to be an average of the three different gases.

\begin{figure*}
    \centering
    \includegraphics[width=\textwidth]{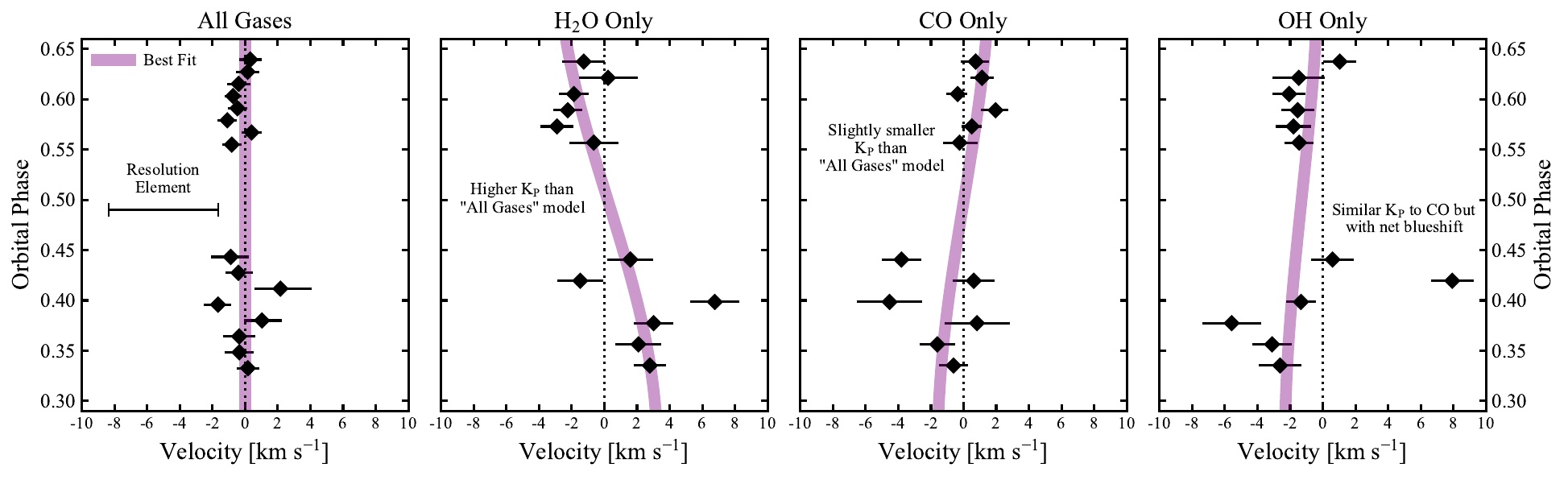}
    \caption{Measured velocities for each gas and the full atmospheric model in the planet rest frame as measured using the ``All Gases" model. The H$_2$O signal has a notable larger $K_P$ compared to the other gases (manifesting here as a greater slant to the left). A possible explanation is that the CO and OH signals are mostly originating from WASP-121 b's hot spot region and are effectively blue- and red-shifted as the planet rotates, resulting in an effectively lower $K_P$ (greater slant to the right). H$_2$O may not experience this effect because it is largely dissociated in the hot spot.}
    \label{fig:gas trails}
\end{figure*}

\new{There is already precedent for measured velocity offsets between these gases in WASP-121 b's atmosphere. Also using IGRINS, \cite{wardenier2024} recently measured offsets between H$_2$O and CO in transmission. The qualitative behavior of both the velocity differences between gases as well as their shifts over the course of transit are best explained by Global Circulation Models (GCM's) with strong atmospheric drag, adding another line of evidence in favor of this scenario as well as demonstrating the ability to probe the three-dimensional nature of this planet using IGRINS data.}

\new{In the context of emission spectroscopy,} the qualitative behavior of these velocity differences between individual gases relative to each other are similar to those observed by \cite{brogi2023} in the atmosphere of UHJ WASP-18 b, also using IGRINS data. Similar to what those authors hypothesize for the case of WASP-18 b, a possible explanation for our measurements is the dissociation of H$_2$O in WASP-121 b's hot spot mitigating the effects of planetary rotation \citep{parmentier2018}. Rotation can cause an apparent blue shift of a molecular signal in the pre-eclipse phases that progressively red shifts as the planet rotates \citep{zhang2017, beltz2022}. This behavior would manifest as an effectively lower $K_P$, which may be why CO and OH, which are still relatively abundant in the the hot spot, have lower measured $K_P$'s compared to H$_2$O. Due to OH's localization to the hot spot and within the hot spot, to intermediate pressures where jet speeds are strongest, we might expect the effects of rotation and jets to be stronger and result in a net red shift rather than the relative blue shift we measure here. However, creating toy models to explain Doppler shifting of molecular signals in emission is nontrivial as there is a complex interplay between the varying temperature and lapse rate with longitude and how each effects the contribution to the total measured signal \citep{vansluijs2023}.

To visualize the anomalous Doppler shifts of each gas as well as their differences between each other, we utilize the method for measuring phase-resolved Doppler shifts described in \cite{pino2022}. This entails optimizing the conditional likelihood of some change in line-of-sight velocity, $\Delta v$, given a best fit orbital velocity solution and placing confidence intervals on $\Delta v$ using Wilk's Theorem. We measure the line-of-sight velocities of H$_2$O, CO, and OH as well as the full atmospheric template in the planet rest frame, for which we use the best fit $K_P$ and $dV_\mathrm{sys}$ as measured using the ``All Gases" model in each sequence. To build signal, we measured these velocities in phase bins of 3 frames for the full model and bins of 6 frames for the individual gases. These phase resolved Doppler shifts are shown in Figure \ref{fig:gas trails} and compared to the best-fit orbital velocity for each model. The measured Doppler shifts for each gas agree well with the path predicted by our measured $K_P$ and $dV_\mathrm{sys}$ values, although the pre-eclipse measurements are much more scattered due to that sequence's lower SNR, and there is a notable outlier in the pre-eclipse OH signal around phase 0.42. This scatter despite relatively small confidence intervals is likely due to unaccounted for correlated noise (such as in time or across wavelength channels within the same telluric line), which is nontrivial to measure even at low spectral resolution.

Dynamical modelling \new{such as that performed in \cite{wardenier2024}} is needed to further support our qualitative and speculative explanations for these measured velocity differences. \new{However, computing and post-processing GCM's as in that study} is beyond the scope of this work. \new{Likewise, \cite{wardenier2024} do not perform atmospheric retrievals as we do in the following section, highlighting both the richness and complexity of high resolution exoplanet spectra}. The difference in measured velocity between gases is rarely more than an IGRINS resolution element (6.67 km s$^{-1}$), so these differences are unlikely to bias retrieved gas abundances as described in the following section. However, again more dynamical modeling coupled with simulated observations and a subsequent retrieval analysis would be required to ascertain any potential biases or lack thereof that arise from atmospheric dynamics in UHJs.

\section{Retrieving Elemental Abundances and the Vertical Thermal Structure}
\label{sec: retrieval}

\subsection{The Forward Model}
\label{sec: forward model}

Using the same likelihood function in Sections \ref{sec: cross correlation} and \ref{sec: dynamics}, we move beyond molecular detection to the estimation of elemental abundances and WASP-121 b's vertical thermal structure via Bayesian inference (``atmospheric retrieval"). We do this with a chemically consistent prescription assuming chemical equilibrium coupled with a flexible P-T profile (both described below). The underlying radiative transfer framework for the forward model is the same GPU-accelerated version of \texttt{CHIMERA}, including opacity sources, used to post-process the 1D-RCTE model in Section \ref{sec: cross correlation}. Each new model spectrum calculated in the posterior sampling process is prepared and filtered as described in that same section.

``Free" chemistry retrievals, in which the VMR of each gas is assumed to be constant with altitude and is individually inferred, are common in the literature (e.g., \citealp{line2013, waldmann2015,molliere2019}). However, as evidenced by our detection of OH and the predictions from the 1D-RCTE model, the composition of WASP-121 b's atmosphere is unlikely to be vertically uniform due to thermal dissociation and ionization of many gases. \cite{brogi2023} recently demonstrated the challenges of using free chemistry models for UHJ atmospheres due to their inability to account for full elemental inventories in the presence of dissociation/ionization. Thus, we opt for a chemically consistent approach using the equilibrium chemistry code GGChem \citep{woitke2018}, which solves both gas phase and condensation equilibrium chemistry and has seen use in retrievals in recent years (e.g., \citealp{zhang2019, alrefaie2022}). Using GGChem, we infer individually each element's enrichment relative to its solar photospheric value from \cite{asplund2009}. The chemical composition of the model atmosphere is thus driven by 9 free parameters: [O/H], [C/H], [Fe/H], [Mg/H], [Si/H], [Ca/H], [Ti/H], [V/H], and [Cr/H]. GGChem then outputs a mean molecular weight profile and the gas phase VMR profiles of each species relevant to the opacity sources listed in Section \ref{sec: cross correlation}, including H, H$^-$, and e$^-$. Hence, H$_2$ is not necessarily a filler gas, especially at low pressures/high temperatures in which it is dissociated.

The chemical composition also depends on the input P-T profile. Our P-T prescription attempts to strike a balance between making few assumptions about the shape of the profile while also keeping the number of model dimensions low. It consists of four pressure ``nodes", one each at the bottom and top of the atmosphere and two additional pressures that can take on any value between. The temperatures at each node can also take on any value, and the temperatures at the four nodes are interpolated onto a finer pressure grid using a B\'ezier spline.

Also included as free parameters in the posterior sampling process are the deviation from the literature radial velocity semiamplitude $dK_P$ and deviation from the total known systemic velocity $dV_\mathrm{sys}$. Unlike in Section \ref{sec: dynamics} or in e.g., \cite{line2021} or \cite{brogi2023}, we do not include the scaling term $a$ because in principle, after filtering the model before each likelihood evaluation as described in Section \ref{sec: cross correlation}, any line amplitude mismatches between the data and the model would be rectified with a more accurate model, and the inclusion of $a$ introduces unnecessary degeneracies. This assumes, of course, that all other relevant system parameters are correct, but any inaccuracies affecting the continuum and/or planet-star contrast are effectively nulled in the detrending process. We use 500 live points and a log(evidence) tolerance of 1 with Pymultinest to sample the posterior distribution. With the radiative transfer calculation accelerated with a NVIDIA A100 GPU and likelihood evaluations parallelized over 12 Intel Broadwell CPU's, the retrieval took $\sim$ a month to complete after 1.3 million likelihood evaluations. The results are described in the following subsection.

\subsection{Retrieval Results}
\label{subsec: retrievals}

\begin{figure*}
    \centering
    \includegraphics[width=0.9\textwidth]{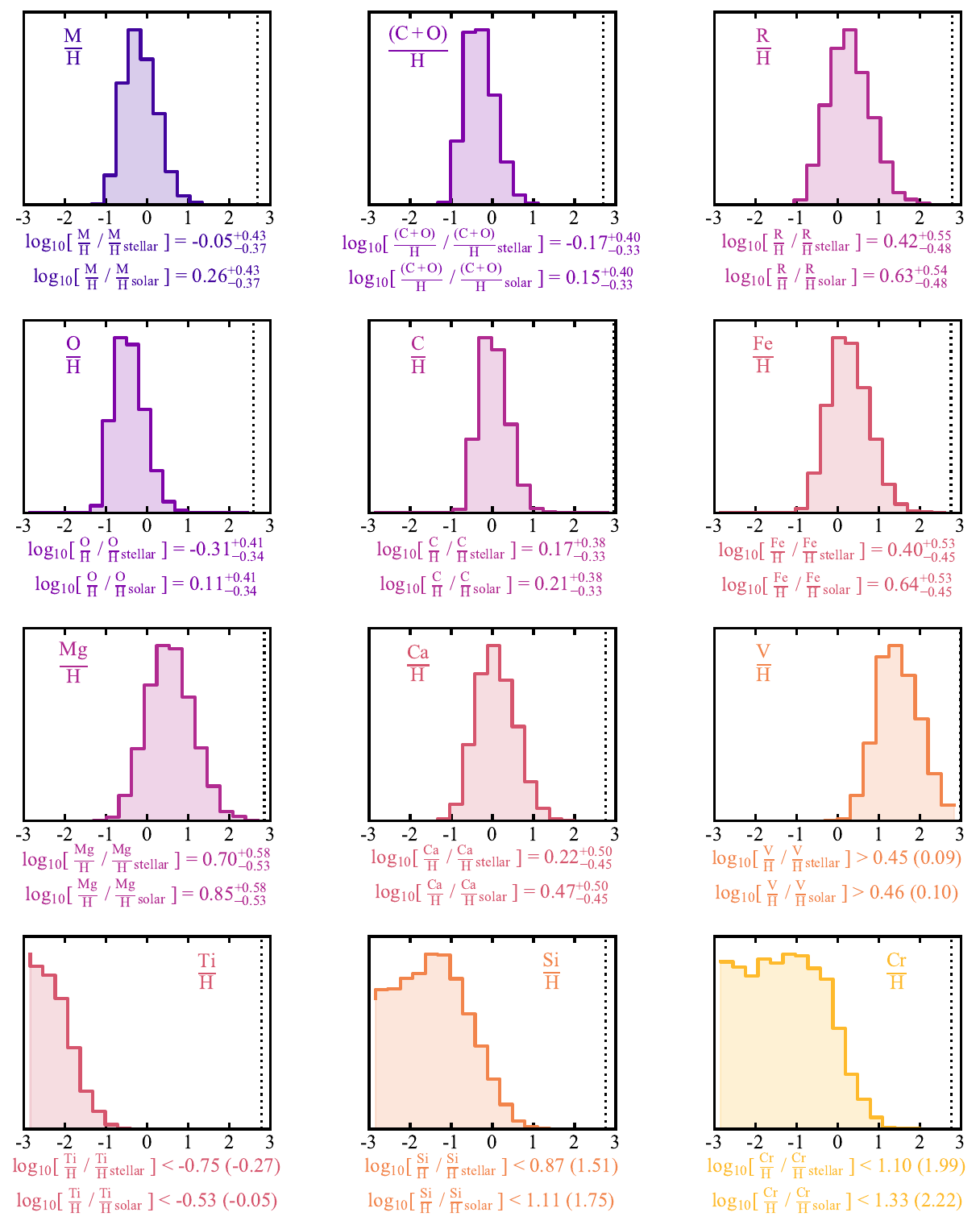}
    \caption{Marginalized posterior distributions for each of the elemental enrichments included in the chemically consistent retrieval model with both volatile and refractory species. When constrained, the posterior median and 1$\sigma$ confidence intervals are listed. For posteriors against the prior boundaries, 3 and 5$\sigma$ upper or lower limits are listed. These quantities were retrieved as log$_{10}$[(x/H) /  (x/H)$_\mathrm{solar}$] and have been converted here to log$_{10}$[(x/H) /  (x/H)$_\mathrm{stellar}$] assuming the values and associated uncertainties for WASP-121 from \cite{polanski2022}. Because WASP-121 has a slightly supersolar metallicity, the upper prior bound on [x/H]$_\mathrm{solar}$ at +3 is slightly lower than 3 when converted to [x/H]$_\mathrm{stellar}$ for any given element. These effective upper prior bounds are indicated by the vertical dotted lines, and the effective lower bounds are less than -3 and beyond the plot. The quantities in the top row are, from left to right, the total atmospheric metal enrichment, the volatile enrichment, and the total refractory enrichment. These are derived from the individual elemental enrichment posterior distributions as described in \new{Section \ref{sec: discussion}}. For reference, the elemental enrichment relative to the solar value is also listed.}
    \label{fig:CC2 panel}
\end{figure*}

\begin{figure*}
    \centering
    \includegraphics[width=0.7\textwidth]{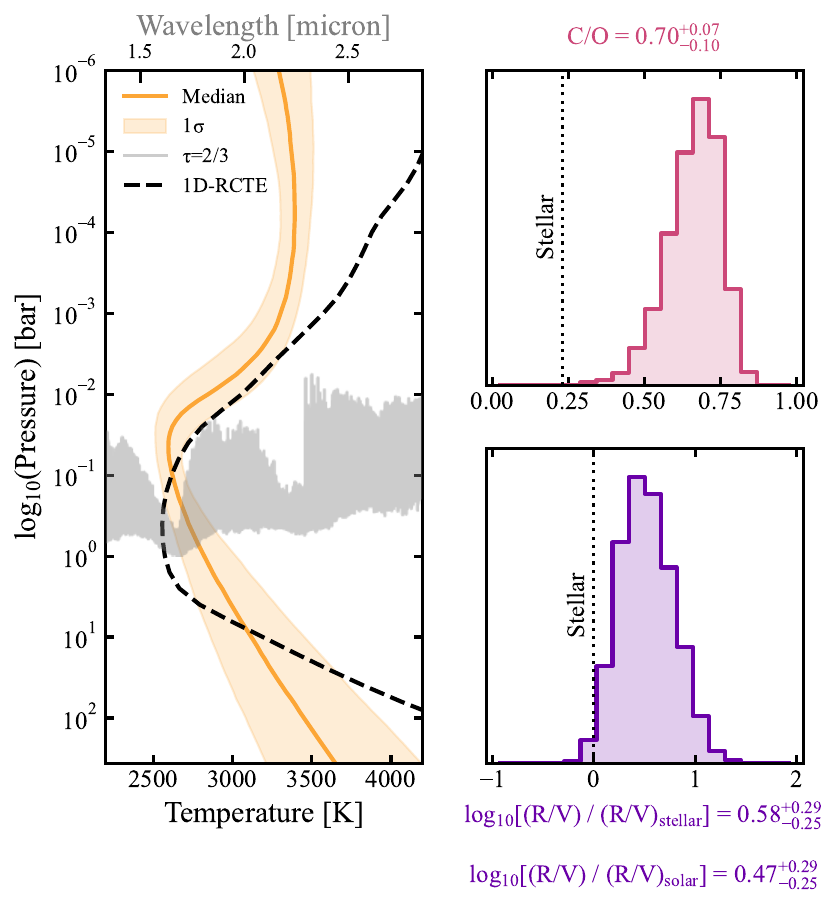}
    \caption{\textit{Left}: The median retrieved P-T profile using the model as described in Section \ref{sec: forward model} (orange line) as well as the 1$\sigma$ confidence interval about this median profile (orange shaded region). For comparison, we also plot the output P-T profile from a 1D-RCTE model at the retrieved best fit metallicity and C/O ratio (black dashed line). Also shown is the $\tau$=2/3 spectrum of the retrieved best fit model, which approximates the location of the photosphere (light gray). \textit{Top right}: The derived posterior distribution for the C/O ratio from that same model. For comparison, the stellar value (0.23) is shown as the dotted black line. \textit{Bottom right}: The derived posterior distribution for the refractory-to-volatile ratio in relation to the stellar value.}
    \label{fig:CC2 summary}
\end{figure*}

The retrieval results are summarized in Figures \ref{fig:CC2 panel} and \ref{fig:CC2 summary}, while the full corner plot can be viewed in the appendix (Figure \ref{fig:corner}). As expected from our CCF detections of H$_2$O, CO, OH, we are able to measure [O/H] and [C/H] to within $\sim$0.4 dex precision. Additionally, we are able to place constraints on the refractory elemental enrichments [Fe/H], [Mg/H], and [Ca/H] to 0.3-0.6 dex precision. We place upper limits on [Ti/H], [Si/H], and [Cr/H] and a lower limit on [V/H].

Out of this selected set of refractory species, it is not surprising that Mg and Ca are the most readily constrained because these two are the most robustly detected refractory species using the log$\mathcal{L}$ maps in Section \ref{sec: cross correlation}. We would also expect them to be among the most abundant refractory species in the pressures IGRINS probes. Fe is similarly abundant at these pressures, but it was only weakly/marginally detected via the log$\mathcal{L}$ map. However, the constraint on [Fe/H] is likely made possible by our sensitivity to individual line strength ratios and the cumulative opacity effect Fe I's strong and broad lines has on these ratios for other species.

Si I is also expected to be abundant at these pressures, but it has very few lines in the H and K bands. Similarly, Cr I has few lines in the H and K bands and the lack of a constraint is not surprising. It is possible that Si and Cr have similar enrichments to the other species, which would be consistent with their upper limits at $\sim20 \times$ solar, but more data are required to determine this. Unlike Si and Cr, the upper limit on the Ti abundance favors a depletion with [Ti/H]$<$ -0.75, which is consistent with previous nondetections of Ti and TiO on WASP-121 b, indicative of a titanium cold trap \citep{merritt2020, hoeijmakers2020, maguire2023, gandhi2023, hoeijmakers2024}.

Of the constrained elemental enrichments, each is solar-to-supersolar, and we derive a total atmospheric metallicity of [M/H]$_\odot$=0.26$^{+0.43}_{-0.37}$. This is done by first converting each elemental solar abundance considered here to the native ratio relative to H and summing these ratios for a total solar M/H fraction. We then do the same with the posterior samples and compare the retrieved M/H to this fiducial solar metal fraction then take the logarithm of the ratio. Due to our retrieved depletion of Ti likely being an effect of cold-trapping, we test whether including it affects the final retrieved [M/H] or refractory content and found no differences. We similarly test whether the inclusion of V affects these derived values and also find negligible differences. Even though V \new{appears to be} significantly enriched (the [V/H] posterior is against the upper prior bound of +3, Figure \ref{fig:CC2 panel}), its contribution to the total number density of metals is relatively small compared to Fe, Mg, and Ca or the volatiles. By taking 1000 random state vector draws from the posterior and passing them through GGChem again, the 3$\sigma$ upper limit on the combined VMRs of V I and VO is $\sim 10^{-5}$, roughly 1\% of the total metal content. Thus, [V/H] is likely not biasing our interpretations of the atmospheric refractory content. 

Likewise, our inability to get a strong detection of Fe via CCF maps calls into question our bounded constraint on [Fe/H]. Again, by excluding [Fe/H] from our metallicity calculation, the differences are negligible. The median retrieved metallicity is 0.03 dex lower, but this does not impact the qualitative interpretation of the metal and refractory content, including the refractory-to-volatile ratio (discussed in the next section).

Splitting the elements into volatile and refactory species and summing them in a similar fashion as above, we derive a roughly solar volatile enrichment at [(C+O)/H]$_\odot$=0.15$^{+0.40}_{-0.33}$ and a moderately supersolar refractory enrichment of [R/H]$_\odot$=0.63$^{+0.54}_{-0.48}$. This results in an overall supersolar refractory-to-volatile ratio of [R/V]$_\odot$=0.47$^{+0.29}_{-0.25}$. The derived C/O ratio is also supersolar at 0.70$^{+0.07}_{-0.10}$. The implications of these two measurements for the planet's formation history are discussed in the following section.

Also as expected per 1D-RCTE modeling and previous works \citep{mikalevans2017, changeat2024, hoeijmakers2024}, we retrieve a thermal inversion layer. To assess the physical plausibility of our retrieved P-T profile, we recomputed a new 1D-RCTE model at the retrieved best fit metallicity and C/O ratio ([M/H]$_\odot$ = -0.18, C/O= 0.75). This model's P-T profile is shown in black in Figure \ref{fig:CC2 summary}, and the retrieved P-T profile shows good agreement with it within the infrared photosphere. While the thermal inversions between the 1D-RCTE profile and our retrieved P-T profile have similar slopes, there is a slight offset in pressure space between the two. Within the context of the inherent degeneracy between metallicity and what pressures the thermal inversion layer occurs, this highlights the inflexibility of the 1D-RCTE model as it only has a single metallicity input value, yet as we show here individual elements can have different enrichments.

We additionally performed a retrieval analysis using a similar chemically consistent prescription but including only the volatile species. We yield consistent inferences on the chemical composition ([(C+O)/H]$_\odot$= -0.02$^{+0.51}_{-0.40}$, C/O=0.65$^{+0.08}_{-0.10}$). However, the inclusion of the refractory species is strongly favored via the Bayes factor at ln($\mathcal{B}$)=73.

\subsection{\new{Refractory-Refractory Ratios and Comparison to Previous Work}}
\label{subsec: comparison}

\begin{figure*}
    \centering
    \includegraphics[width=\linewidth]{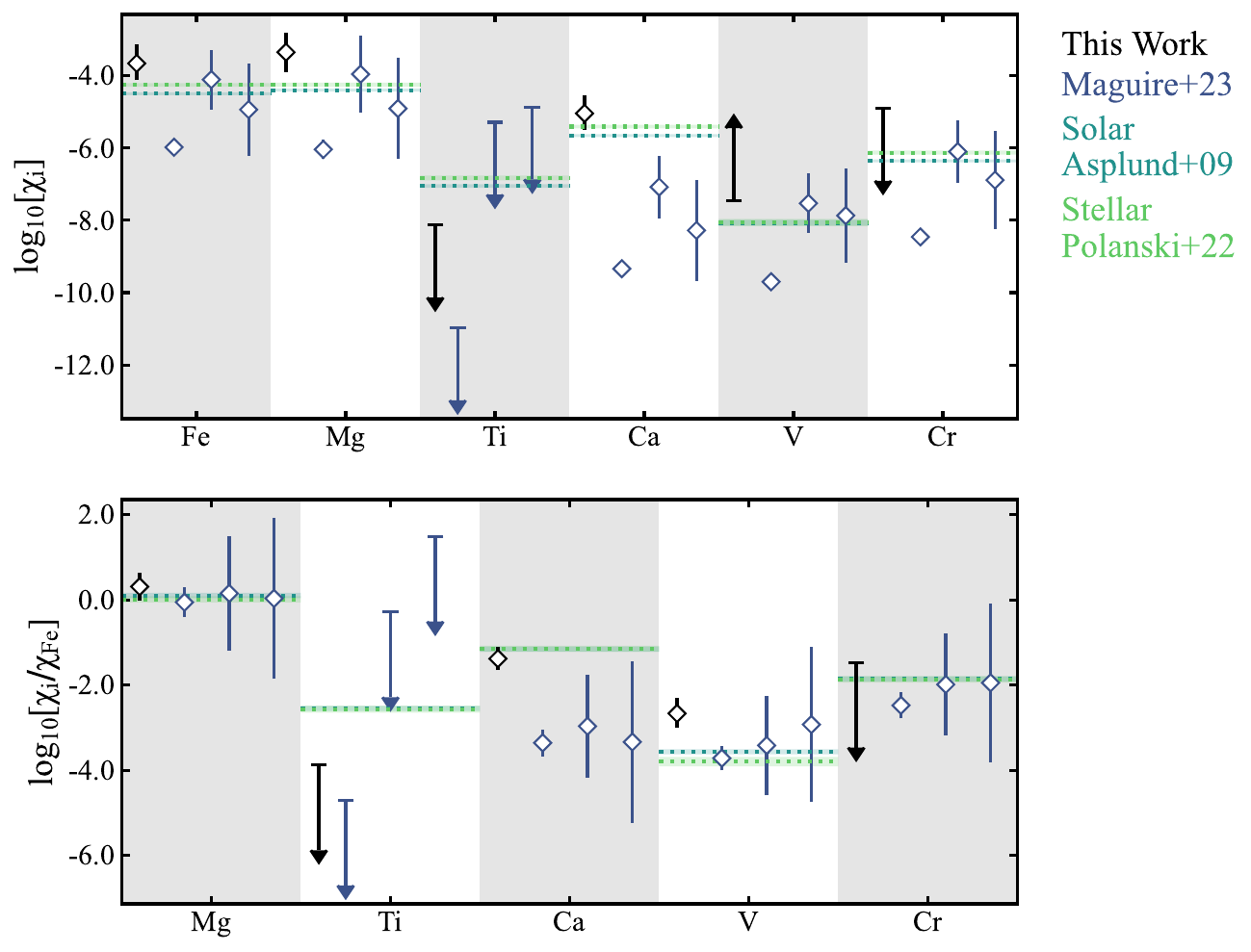}
    \caption{\new{Our measured absolute abundances (\textit{top panel}) and relative abundances (to Fe, \textit{bottom}) in black compared to those measured by \cite{maguire2023} over the course of 3 transits (dark blue). Also shown are the solar \citep{asplund2009} and stellar \citep{polanski2022} photospheric values for these quantities (turquoise and green horizontal lines, respectively). While there is some scatter among the absolute abundances, the relative abundance ratios are mostly consistent across these two studies with the exception of Ca. As discussed in Section \ref{subsec: comparison}, this may be an artifact of the different chemistry prescriptions used in each study, or this may be evidence of partial Ca condensation on the planet's terminator. Arrows represent 3$\sigma$ upper or lower limits.}}
    \label{fig:maguire}
\end{figure*}

\new{This study marks one of the first simultaneous measurements of multiple refractory elemental abundances in a transiting exoplanet via high resolution spectroscopy in the near infrared. To interrogate the plausibility of our measurements, we compare to those from \cite{maguire2023}, which is one of the most comprehensive retrieval studies of WASP-121 b at high spectral resolution. From three transits observed with ESPRESSO/VLT, they measure the absolute abundances of several species also considered in this study: Fe I, Mg I, Ti I, V I, Ca I, and Cr I. However unlike here, they use a ``free" chemistry modeling framework that assumes constant-with-altitude gas VMR's. To compare observable quantities, we post-process our retrieval posterior distribution of elemental enrichments through GGChem to obtain a posterior distribution of pressure-dependent gas VMR profiles. We then find the photospheric average abundance of each gas mentioned above using the wavelength integrated contribution function of the best fit model spectrum. The median of these photospheric abundances and 1$\sigma$ confidence intervals are shown in the top panel of Figure \ref{fig:maguire} alongside these values as measured by \cite{maguire2023}.}

\new{While there is some scatter in these absolute abundance measurements (both between our measurements and those of \cite{maguire2023} and among the different transits presented in that study), \cite{maguire2023} aptly highlight the difficulty in continuum normalization for high resolution spectroscopy. This is especially relevant in the case of transmission spectroscopy, in which the degeneracies between parameteters controlling the continnuum -- such as planet white-light radius, cloud deck pressure, and temperature -- are difficult to break. This weakness is compensated by a high sensitivity to abundance \textit{ratios}. Thus, we also compare each of our measured gas abundance ratios compared to Fe I to the same such measured ratios from \cite{maguire2023}.} 

\new{As can be seen in the bottom panel of Figure \ref{fig:maguire}, these refractory-refractory ratios are overall consistent both with the \cite{maguire2023} measurements as well as the stellar photospheric elemental ratios from \cite{polanski2022} (with the exception of Ti). Most inconsistent with \cite{maguire2023} is our measurement of the Ca/Fe abundance ratio. However, our measurement of this ratio is consistent with the stellar value, which is in line with expectations that unlike the volatiles, refractory elemental ratios throughout the protoplanetary disk should not vary from the stellar ratios.}

\new{The source of this discrepancy with \cite{maguire2023} is unclear. However, \cite{gandhi2023} notably also measure a substellar Ca abundance and Ca/Fe abundance ratio for this planet via high resolution transmission spectroscopy. Ca has a similar condensation temperature to Ti \citep{lodders2003,wakeford2017}, which all evidence points to being globally cold-trapped. This condensation temperature is near the 1550 K cold trap transition proposed by \cite{pelletier2023}, and even if Ca is not being sequestered into Ti-bearing condensates, there are numerous other Ca-bearing condensates with condensation temperatures near or above the Ti condensation temperatures (e.g., hibonite, grossite, and gehlenite). Thus, it is likely that Ca is abundant on the planet day-side, and the ESPRESSO transmission observations are probing longitudes at which Ca is being partially depleted via condensation before transitioning to fully condensed out on the night side.}

\new{We note the difficulty in comparing retrieved abundances using different chemical modeling prescriptions. Both \cite{maguire2023} and \cite{gandhi2023} assume constant-with-altitude VMRs. However, both our posterior VMR profile draws and 1D-RCTE modeling indicate that, at least on the day side, Ca I ionizes at the upper edge of the photosphere and at lower altitudes than Fe I. Thus, free chemistry retrievals may be biased toward lower Ca abundances and lower Ca/Fe ratios regardless of any partial condensation. Our method of photospheric averaging should have reproduced such a bias as measured via near infrared emission spectroscopy, but optical transmission spectra are likely probing higher altitudes and thus more sensitive to the ionization of Ca I. Combining both the emission and transmission data sets into a uniform retrieval process would help alleviate these discrepancies and work towards confirming this global ``calcium cycle", but this is left for future work.}

\new{Slightly disparate from the stellar abundance ratio is our V/Fe measurement. While it is consistent with two of the \cite{maguire2023} transits, our measurement is 3$\sigma$ greater than the stellar value and \cite{maguire2023}'s most precise measurement. This it likely due to the reliance of our [V/H] inference on both atomic V I and the oxide VO. In particular, the \cite{mckemmish2016} VO line list has been shown to be incomplete, coincidentally in the context of analyzing high resolution spectra of WASP-121 b \citep{deregt2022}. That study demonstrated that even with injected VO signals at the abundance inferred by \cite{evans2018}, VO would still not be recovered. Thus, we are likely overestimating the total V inventory. During the preparation of this manuscript, the ExoMol VO line list was updated \citep{bowesman2024, mckemmish2024}, and a reanalysis is warranted. However, as discussed above, V I and VO make up only a small fraction of the total metal content in WASP-121 b's atmosphere and our qualitative interpretations are likely not significantly biased. Thus we save such a reanalysis for future studies.}

\section{Discussion and Implications for Planet Formation}
\label{sec: discussion}

\begin{figure*}
    \centering
    \includegraphics[width=0.65\textwidth]{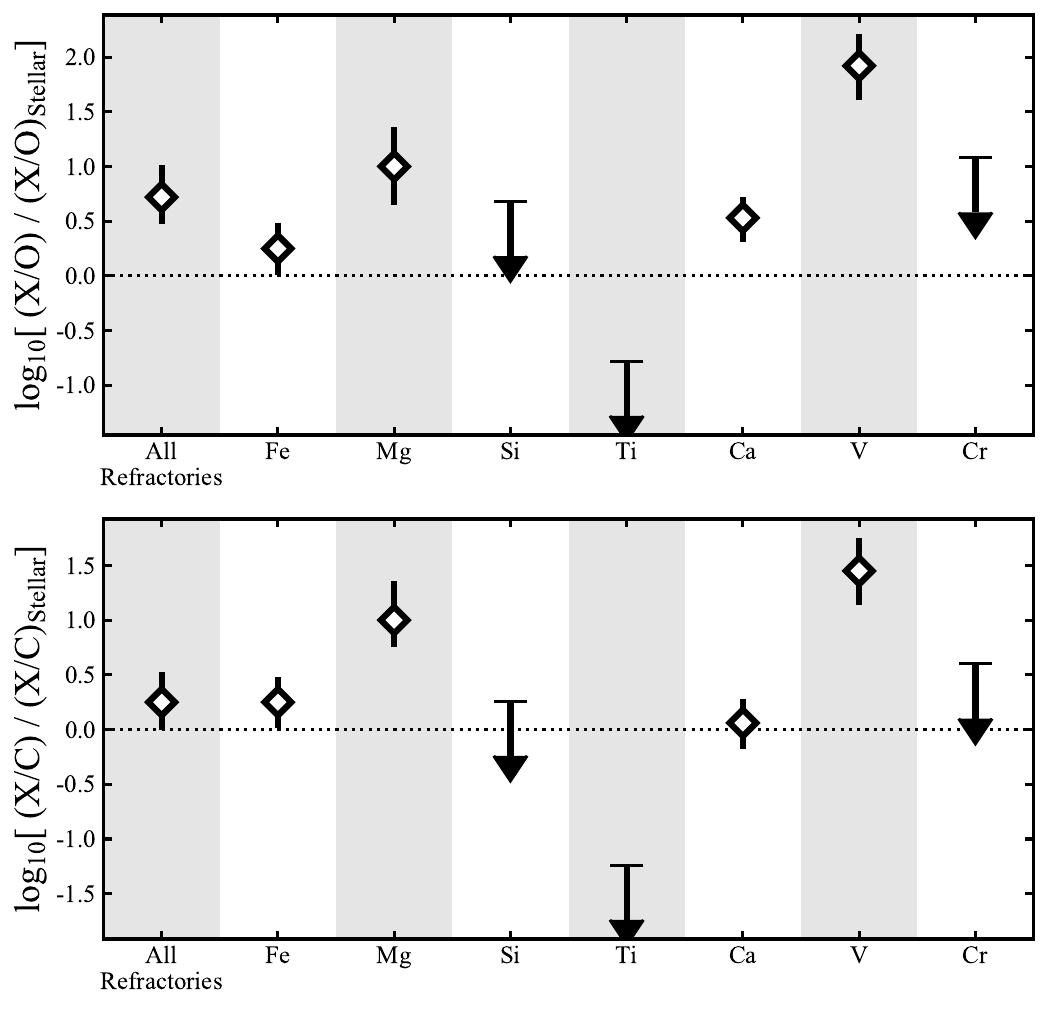}
    \caption{Abundance ratios of refractory species both in total and individually to oxygen (top) and carbon (bottom) in terms of the stellar ratios. Arrows indicate upper limits at 3$\sigma$.}
    \label{fig:ratios}
\end{figure*}

\begin{figure*}
    \centering
    \includegraphics[width=\textwidth]{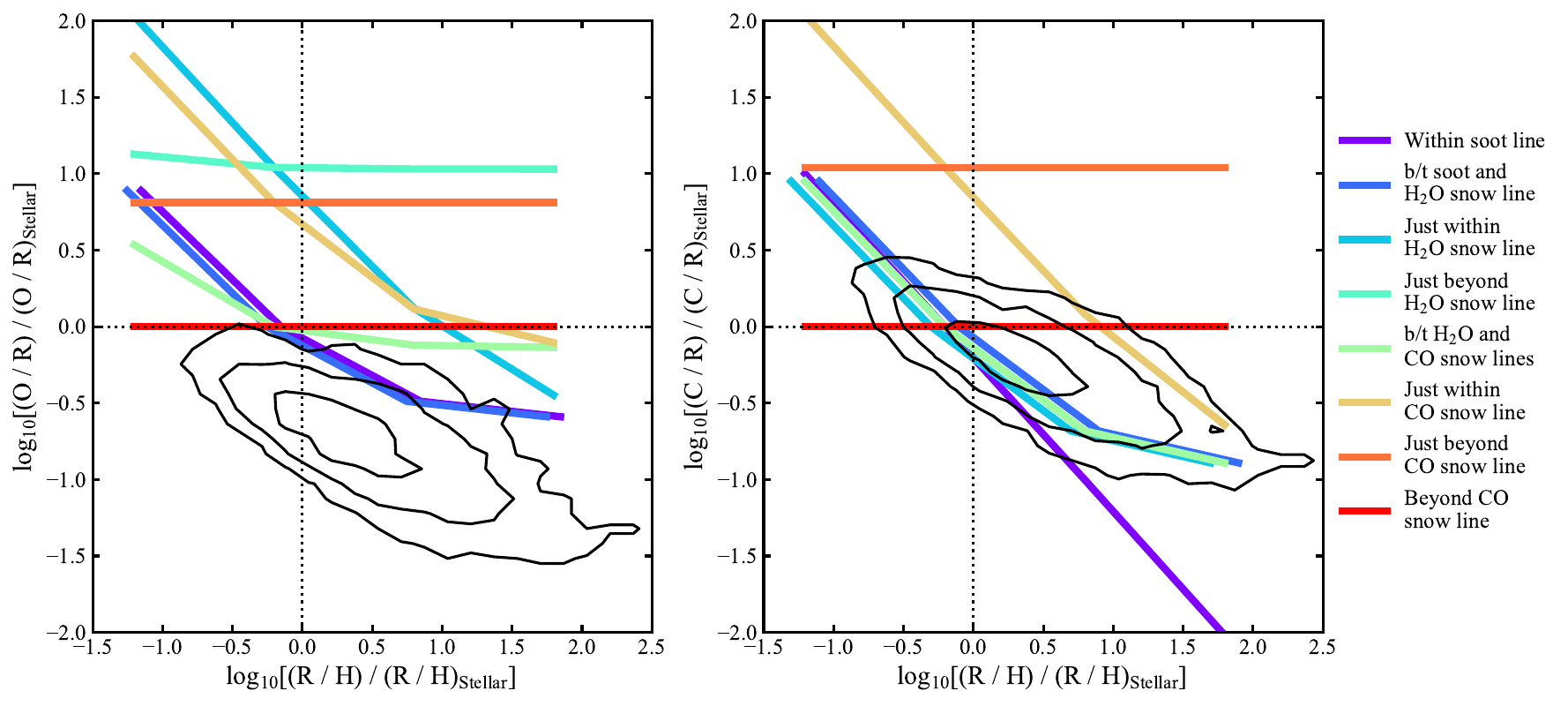}
    \caption{Our measurements for WASP-121 b's atmospheric oxygen- and carbon-to-refractory ratios and overall refractory enrichment using the model described in Section \ref{subsec: retrievals}, compared to \cite{chachan2023}'s predictions for these values depending on initial formation location. Some of these curves have been horizontally offset by up to 0.1 dex for visual clarity. The black contours are the 1, 2, and 3$\sigma$ quantiles for these 2D posterior distributions.}
    \label{fig:chachan predictions}
\end{figure*}

The combination of metallicity, C/O, and refractory-to-volatile (R/V) ratio measured using a single instrument provides robust diagnostic power for inferring the planet's formation history. Thus far we have discussed WASP-121 b's atmospheric composition in relation to solar composition, but comparison to the host star's composition is more informative for inferring the planet's formation history. Using the host star chemical abundances\footnote{Stellar abundances are nontrivial to determine and can be updated frequently (see e.g., \cite{fortney2012} and \cite{bedell2018}). In the event that the WASP-121 stellar abundances change in the future, it is straightforward to update our retrieved elemental abundances for the planet by adding the new stellar values to the listed value relative to solar in Figure \ref{fig:CC2 panel}.} as well as the associated error bars\footnote{We subtract from each posterior random values from a normal distribution centered at the stellar abundance and with a standard deviation equal to the uncertainties reported in Table \ref{tab:params}.} from \cite{polanski2022}, our inferred planet atmospheric composition is consistent with a stellar metallicity and a superstellar C/O ratio, with our retrieved metallicity and C/O ratio at [M/H]$_\star$=-0.05$^{+0.43}_{-0.37}$ and C/O=0.70$^{+0.07}_{-0.10}$ (3.03$^{+0.32}_{-0.42}$ $\times$ stellar, which is 0.23$\pm$0.05).

With a near-Jupiter mass ($M_P$=1.183$^{+0.064}_{-0.062} M_\mathrm{Jup}$, \citealp{delrez2016}), a stellar metallicity for WASP-121 b is not unexpected within the context of mass-metallicity trends measured both in the solar system \citep{atreya2016} and the general transiting exoplanet population \citep{thorngren2019a}. Following the interior-structure derived trend from \cite{thorngren2016} and \cite{thorngren2019a}, WASP-121 b's atmospheric metallicity is consistent with roughly 95\% of the planet's bulk metal content being in the planet core, which should then be $\sim$50 $M_\oplus$. This is in line with \cite{bloot2023}, who infer a similar core mass for WASP-121 b via interior modelling, albeit with large confidence intervals on this value (43$^{+91}_{-30} M_\oplus$). WASP-121 b's carbon enrichment also very closely follows the solar system [C/H] trend \citep{atreya2016}, possibly indicative of a similar formation process between UHJs and the solar system giants, but only so much can be extrapolated from one data point and more UHJ carbon inventory measurements are needed to draw further conclusions on this.

Such mass-metallicity trends are a distinctive prediction of the core accretion paradigm of planet formation. Within the core accretion paradigm, the combination of a stellar metallicity and a significantly elevated C/O ratio compared to the stellar host (C/O=0.23$\pm$0.05) can be indicative of formation exterior to the H$_2$O snow line and subsequent migration through the protoplanetary disk \citep{oberg2011, mordasini2016, cridland2019}. This is one hypothesis for the origin of hot and ultra-hot Jupiters \citep{lin1996}, although high eccentricity migration is also needed to explain the observed population \citep{fortney2021}. Formation beyond the snow line and later migration is the same conclusion drawn by \cite{changeat2024}, who also infer WASP-121 b to have a subsolar-to-solar metallicty \new{(-0.77 $<$ log($Z$) $<$ 0.05)} and supersolar C/O ratio \new{(0.59 $<$ C/O $<$ 0.87)} via HST phase curve measurements.

However, measurements of the C/O ratio alone can be degenerate with planet formation pathways \citep{mordasini2016, schneider2021, turrini2021}, and measurements of WASP-121 b's atmospheric refractory content can help break such degeneracies. Compared to the stellar photospheric composition, WASP-121 b's volatile content is roughly stellar at [(C+O)/H]$_\star$= -0.17$^{+0.40}_{-0.33}$ (0.67$^{+1.00}_{-0.36} \times$ stellar), while its refractory content is stellar-to-superstellar at [R/H]$_\star$=0.42$^{+0.54}_{-0.48}$ (2.64$^{+6.00}_{-1.77} \times$ stellar). From these two, the derived refractory-to-volatile ratio (R/V), is moderately superstellar at [R/V]$_\star$=0.58$^{+0.29}_{-0.25}$ (3.83$^{+3.62}_{-1.67} \times$ stellar) at 2.3$\sigma$ ($p = 0.02$ via a chi-square survival function). 

The ratios of the total refractory content to C or O individually, as well as individual refractory elements to C and O, are also largely superstellar with the notable but expected exception of Ti (Figure \ref{fig:ratios}). This illustrates that our inference of a superstellar R/V ratio is not skewed by any one element. An elevated R/V ratio indicates enhanced solid accretion, possibly via pollution by planetesimals that are ``rocky" rather than ``icy", as might be expected beyond the snow line \citep{lothringer2021, schneider2021}. This is consistent with the findings of \cite{lothringer2021}, who measure WASP-121 b's atmospheric refractory-to-oxygen ratio to be [R/O]$_\odot$=0.70$^{+0.34}_{-0.33}$ via HST STIS and WFC3 transmission spectra \citep{evans2018, sing2019}, compared to our measured [R/O]$_\odot$=0.51$^{+0.29}_{-0.25}$.

Within a disk modeling framework, \cite{chachan2023} make several predictions on the final atmospheric R/H, R/O, and R/C based on initial formation location. They state that \cite{lothringer2021}'s measurements are most consistent with formation between the soot line and H$_2$O snow lines. However, they note that without measurements of CO, and thus with incomplete knowledge of the atmospheric O and C inventories, other scenarios are still possible. Comparing our measurements of these quantities with \cite{chachan2023}'s predictions scaled to WASP-121's composition, we can rule out formation interior to the soot line (which would result in a low C/O ratio) or near the CO snow line (low R/O and R/C).

A combination of an elevated R/V ratio (indicating refractory-rich planetesimal pollution interior to the H$_2$O snow line) and high C/O ratio (placing formation outside of the carbon-depleted region interior to the soot line) is most consistent with formation between the soot line and H$_2$O snow line (Figure \ref{fig:chachan predictions}), in line with \cite{chachan2023}'s previous interpretation of \cite{lothringer2021}'s measurements. However, it should be noted that WASP-121 b's R/O and R/C ratios could both be stellar within 2.6$\sigma$ ($p$=0.01). In this case, formation beyond the CO snow line would be possible regardless of the total refractory enrichment, but this would require inward migration of 10's of au compared to $\lesssim$3 in the other scenario. Additionally, with our measured super-stellar C/O ratio, formation between the H$_2$O and CO snow lines would be possible only if the planet's final refractory enrichment is substellar, but we cannot rule this out at 2.5$\sigma$ ($p$=0.01 for [R/H]= -1). More observations would likely enable us to place more stringent constraints on these quantities, but even with highly precise measurements of a planet's atmospheric composition, unambiguously inferring a planet's formation history is nontrivial. E.g., \cite{chachan2023}'s models do not account for migration between regions while the planet is still accreting its atmosphere. The planet may then inherit material from multiple chemically distinct regions of the protoplanetary disk, which itself also evolves in time \citep{molliere2022}.

Formation of gas giant planets interior to the H$_2$O snow line is not a common prediction of the typical ``solar nebula" and core accretion models of planet formation. This is because it is believed the that condensation of ices is required to provide the requisite solid material to form a core massive enough to undergo runaway gas accretion. However, the possibility of formation interior to the snow line has been explored in the literature (e.g., \citealp{bodenheimer2000,lee2014,batygin2016, bailey2018}), and it can be possible under certain conditions. A common finding among these studies is the possibility of the most massive super-Earths initiating runaway gas accretion and eventually becoming gas giants if accretion continues for a requisite amount of time before the disk dissipates. \cite{bailey2018} point out that super-Earths are quite common in the Milky Way, and only a small number ($\sim$1\%) would need to enter this runaway accretion regime to explain the observed population of hot and ultra-hot Jupiters. Ultimately, formation simulations tailored to the case of WASP-121 b would be required to robustly explore the plausibility of formation interior to the snow line (such as, e.g., \citealp{bitsch2022} and \citealp{khorshid2023} in the case of WASP-77A b), but such simulations are beyond the scope of this paper.

\section{Summary and Conclusions}
\label{sec:conclusions}

In this paper, we present observations of the ultra-hot Jupiter WASP-121 b via high resolution emission spectroscopy in the near infrared. Using the IGRINS instrument on Gemini South, which has simultaneous H and K band wavelength coverage at R$\approx$45,000, we captured the direct thermal emission of WASP-121 b in two observational sequences covering the pre- and post-eclipse phases of the planet's orbit. Using standard cross-correlation techniques, we detect the planet's atmosphere at a signal to noise ratio of 8.31. The spectral template used to make this detection is computed from a 1D radiative-convective-thermoequilibrium (1D-RCTE) atmospheric model that includes a thermal inversion layer and thermal dissociation of H$_2$O. Thus, our detection of emission lines rather than absorption lines in the planet's outgoing thermal emission spectrum confirms previous detections of a thermal inversion in WASP-121 b's atmosphere. Searching for individual gases via cross-correlation, we also detect CO, OH, and a weak H$_2$O signal. The latter two are indicative of thermal dissociation of H$_2$O. Using the log-likelihood function from \cite{brogi2019}, we also tentatively detect the individual refractory species Mg I, Ca I, and V I. 

Implementing the log-likelihood function into the nested sampler Pymultinest, we measure WASP-121 b's orbital velocity, $K_P$, to be consistent with literature expectations of a circular orbit. However, we also measure a small net red shift of 1.20$^{+0.13}_{-0.11}$ km s$^{-1}$ when comparing a model spectrum including the opacities of numerous gases to the data. When repeating the inference of $K_P$ and a net $dV_\mathrm{sys}$ using model spectra containing only individual gases, we find that H$_2$O, CO, and OH have slight velocity offsets. This could be indicative of our ability to probe thermochemical inhomogeneities on WASP-121 b's day side, but more work on dynamically modeling the atmosphere is needed to robustly interpret our measurements.

To infer WASP-121 b's atmospheric composition and thermal structure, we apply a chemically consistent model prescription in an atmospheric retrieval analysis. We infer the enrichments of O, C, and multiple refractory elements and are able to measure WASP-121 b to have both a super-stellar C/O ratio and a super-stellar refractory-to-volatile ratio, consistent with previous studies with space-based observatories. \new{Our inferred individual refractory-refractory abundance ratios are also comparable both in precision and value to previous transit observations with ESPRESSO/VLT. Within expectations for disk chemistry, these abundance ratios are consistent with the stellar values with the exception of Ti, which is likely cold-trapped, and V, whose inferred abundance is likely biased by an incomplete VO line list. Notably, our inference of a stellar Ca/Fe abundance ratio departs from previous studies in transmission that measured Ca to be depleted \citep{maguire2023, gandhi2023}. This is possibly indicative of partial Ca condensation on the planet's terminator, but biases due to modeling assumptions in those works could also explain this discrepancy.} 

Comparing to previous disk chemistry modeling efforts, our measured composition is most consistent with formation between the soot line and H$_2$O snow line, which is not a commonly expected formation pathway for giant planets. However, we cannot rule out formation between the H$_2$O and CO snow lines or beyond the CO snow line. Regardless of WASP-121 b's exact formation history, building a large sample size of measured refractory-to-volatile ratios of ultra-hot Jupiter atmospheres will be a crucial step forward in refining our knowledge of how giant planets form. Previous such measurements have been made through joint analyses of data from different instruments, but here we have demonstrated the ability to do so with a single high resolution infrared spectrograph.

\facilities{Gemini South (IGRINS)}

\software{maptlotlib \citep{hunter2007}, numpy \citep{vanderWalt2011}, pymultinest \citep{buchner2016}, scipy \citep{virtanen2019}}

\section*{Acknowledgements}

Arizona State University is located on the traditional and unceded homelands of Indigenous peoples, including the Akimel O’odham (Pima) and Pee Posh (Maricopa) Indian Communities. \new{We thank C. Maguire for helpful discussion and the anonymous referee for their suggestions to improve this manuscript.} P.C.B.S. acknowledges support provided by NASA through the NASA FINESST grant 80NSSC22K1598. M.R.L. and J.L.B. acknowledge support from NASA XRP grant 80NSSC19K0293 and NSF grant AST-2307177. E.R. acknowledges support from the Heising-Simons Foundation. M.W.M. acknowledges support through the NASA Hubble Fellowship grant HST-HF2-51485.001-A. J.P.W. acknowledges support from the Trottier Family Foundation via the Trottier Postdoctoral Fellowship. S.P. acknowledges funding from the NCCR PlanetS supported by the Swiss National Science Foundation under grant 51NF40\_205606. Based on observations obtained at the international Gemini Observatory, a program of NSF’s NOIRLab, which is managed by the Association of Universities for Research in Astronomy (AURA) under a cooperative agreement with the National Science Foundation on behalf of the Gemini Observatory partnership: the National Science Foundation (United States), National Research Council (Canada), Agencia Nacional de Investigaci\'{o}n y Desarrollo (Chile), Ministerio de Ciencia, Tecnolog\'{i}a e Innovaci\'{o}n (Argentina), Minist\'{e}rio da Ci\^{e}ncia, Tecnologia, Inova\c{c}\~{o}es e Comunica\c{c}\~{o}es (Brazil), and Korea Astronomy and Space Science Institute (Republic of Korea). This work used the Immersion Grating Infrared Spectrometer (IGRINS) that was developed under a collaboration between the University of Texas at Austin and the Korea Astronomy and Space Science Institute (KASI) with the financial support of the US National Science Foundation 27 under grants AST-1229522 and AST-1702267, of the University of Texas at Austin, and of the Korean GMT Project of KASI.

\bibliography{main.bib}{}
\bibliographystyle{aasjournal}

\appendix

\section{Traditional CCF detection maps}
\label{A: ccfmaps}

\begin{figure*}[!ht]
    \centering
    \includegraphics[width=\textwidth]{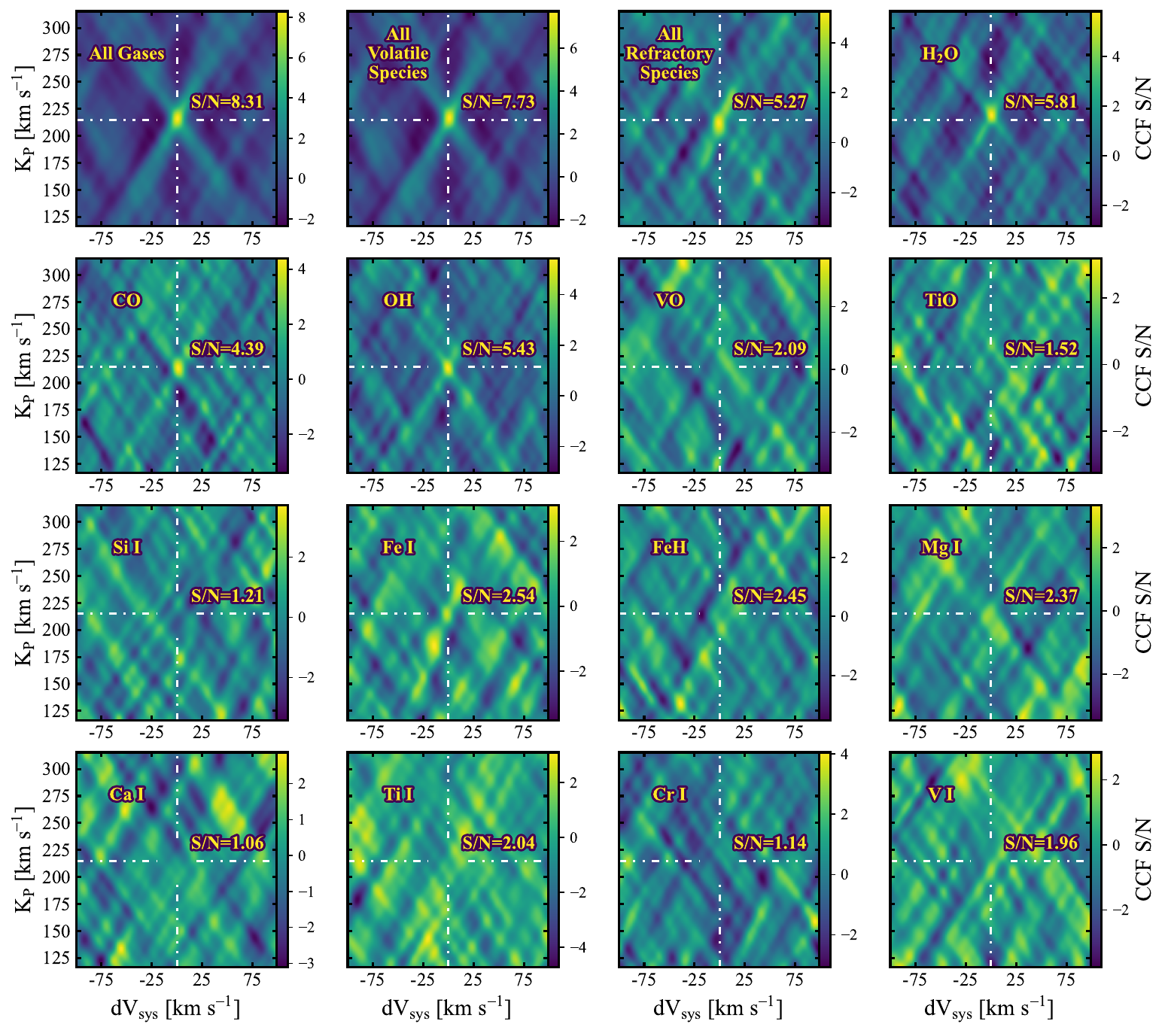}
    \caption{Cross-correlation detection maps calculated by cross-correlating with the data a solar composition model spectrum with only each individual gas included, rather than the residual method as described in Section \ref{sec: cross correlation}. Compare to Figure \ref{fig:CCF maps}}
    \label{fig:vanilla ccf}
\end{figure*}

\begin{figure*}[!ht]
    \centering
    \includegraphics[width=\textwidth]{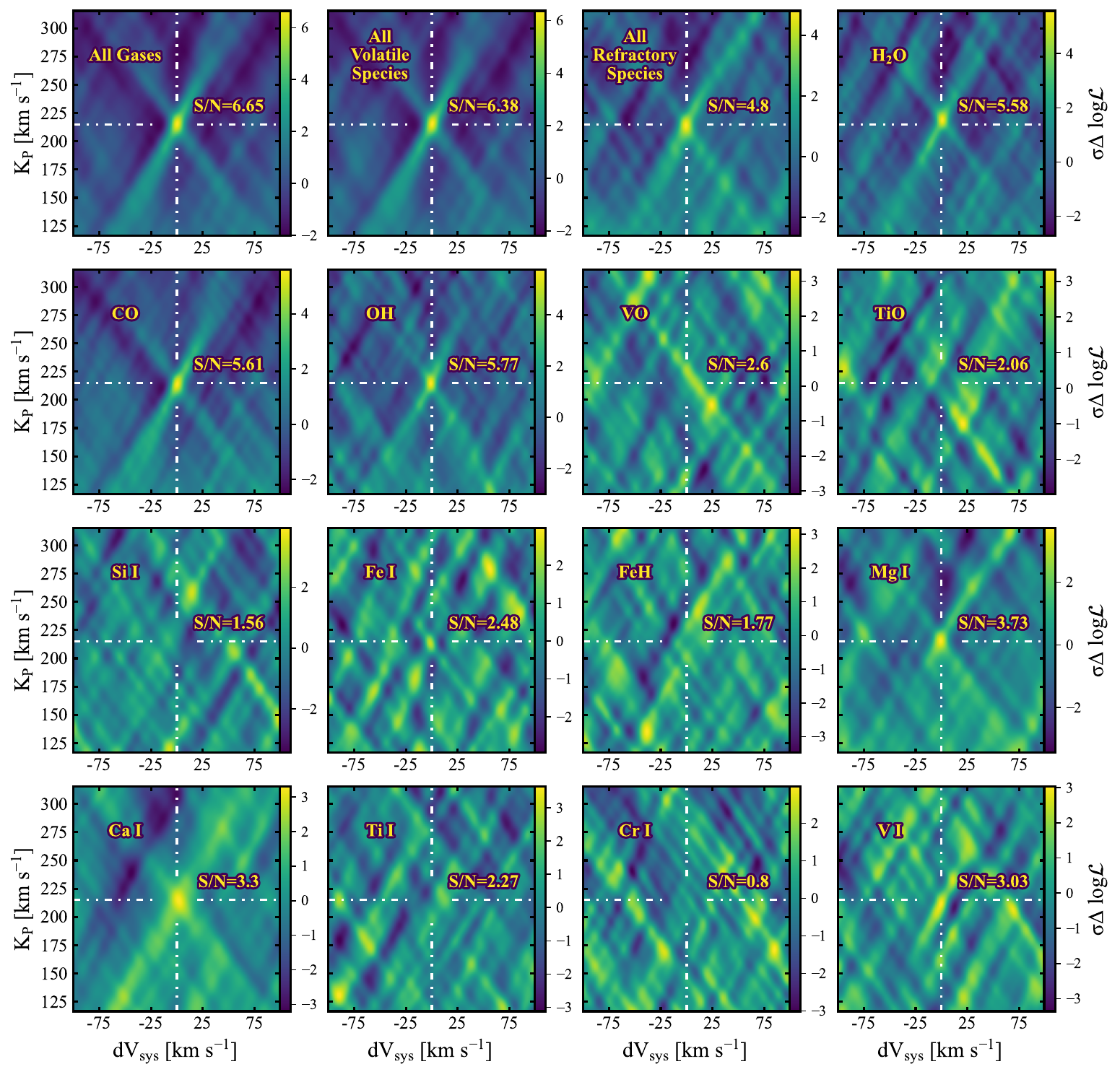}
    \caption{Similar to figure \ref{fig:vanilla ccf}, but log-likelihood detection maps calculated using model spectra including the opacity of only each individual gas. Compare to Figure \ref{fig:logL maps}.}
    \label{fig:vanilla logL}
\end{figure*}

\section{Retrieval Corner Plot}
\label{B: corner}

\begin{figure*}[!ht]
    \centering
    \includegraphics[width=\textwidth]{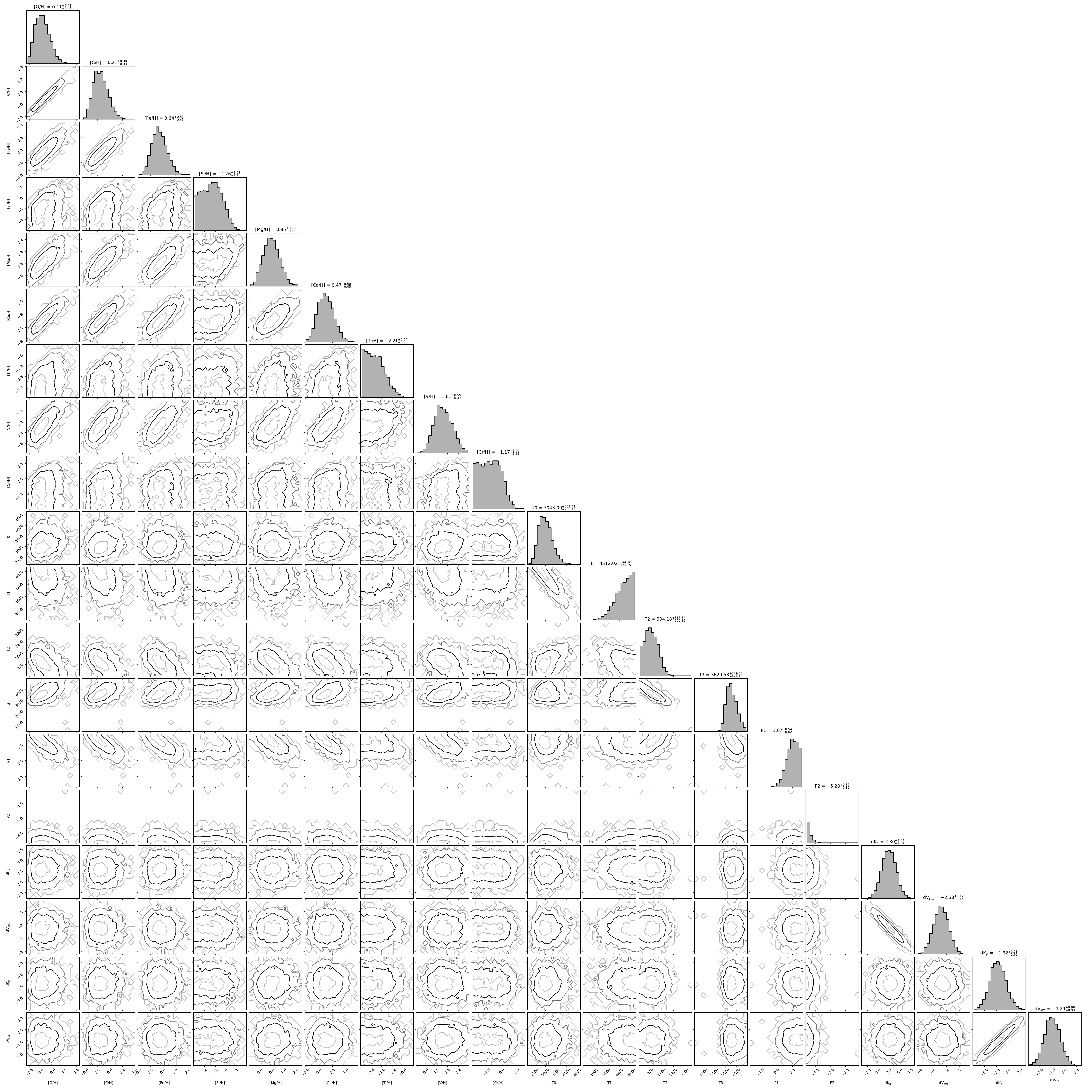}
    \caption{Associated corner plot for the atmospheric retrieval discussed in Section \ref{sec: retrieval}.}
    \label{fig:corner}
\end{figure*}

%TC:endignore

\end{document}